\documentclass[conference]{IEEEtran}
\usepackage{cite}
\usepackage{amsmath,amssymb,amsfonts}
\usepackage{algorithmic}
\usepackage{graphicx}
\usepackage{textcomp}
\usepackage{xcolor}
\usepackage[hyphens]{url}
\usepackage{fancyhdr}
\usepackage{hyperref}
\usepackage{array}
\usepackage{tabularx}
\usepackage{bm}
\usepackage{booktabs}
\usepackage{authblk}

\def\BibTeX{{\rm B\kern-.05em{\sc i\kern-.025em b}\kern-.08em
    T\kern-.1667em\lower.7ex\hbox{E}\kern-.125emX}}

\usepackage{circledsteps}
\newcommand\myCircled[2][]{\ifmmode%
\Circled[fill color=black,inner color=white,#1]{\footnotesize\mathsf{#2}}%
\else%
\Circled[fill color=black,inner color=white,#1]{\footnotesize\sffamily#2}%
\fi%
}

\title{HyperSense: Hyperdimensional Intelligent Sensing for Energy-Efficient Sparse Data Processing}
\author[1, *]{Sanggeon Yun}
\author[1, *]{Hanning Chen}
\author[1]{Ryozo Masukawa}
\author[1]{Hamza Errahmouni Barkam}
\author[1]{\authorcr Andrew Ding}
\author[1]{Wenjun Huang}
\author[1]{Arghavan Rezvani}
\author[2]{Shaahin Angizi}
\author[1, $\dag$]{Mohsen Imani}
\affil[1]{University of California, Irvine}
\affil[2]{New Jersey Institute of Technology}
\affil[*]{Equal contributions.}
\affil[$\dag$]{Corresponding author, email: m.imani@uci.edu}

\begin{document}
\maketitle
\thispagestyle{plain}
\pagestyle{plain}

\begin{abstract}
Introducing \emph{HyperSense}, our co-designed hardware and software system efficiently controls Analog-to-Digital Converter (ADC) modules' data generation rate based on object presence predictions in sensor data. Addressing challenges posed by escalating sensor quantities and data rates, \emph{HyperSense} reduces redundant digital data using energy-efficient low-precision ADC, diminishing machine learning system costs. Leveraging neurally-inspired HyperDimensional Computing (HDC), \emph{HyperSense} analyzes real-time raw low-precision sensor data, offering advantages in handling noise, memory-centricity, and real-time learning. Our proposed \emph{HyperSense} model combines high-performance software for object detection with real-time hardware prediction, introducing the novel concept of \emph{Intelligent Sensor Control}. Comprehensive software and hardware evaluations demonstrate our solution's superior performance, evidenced by the highest Area Under the Curve (AUC) and sharpest Receiver Operating Characteristic (ROC) curve among lightweight models. Hardware-wise, our FPGA-based domain-specific accelerator tailored for \emph{HyperSense} achieves a 5.6$\times$ speedup compared to YOLOv4 on NVIDIA Jetson Orin while showing up to 92.1\% energy saving compared to the conventional system. These results underscore \emph{HyperSense}'s effectiveness and efficiency, positioning it as a promising solution for intelligent sensing and real-time data processing across diverse applications.

\end{abstract}

\section{Introduction}

Ubiquitous sensors, witnessing exponential growth in numbers and data generation rates, pose formidable challenges for existing processing methods due to algorithmic and architectural limitations. In the Internet of Things (IoT), the use of machine learning algorithms for sensor data analysis often leads to compatibility issues, given the escalating data volume requirements. These challenges stem from sensor computing demands, redundant information transmission, and computationally intensive analyses, resulting in time and energy inefficiencies.

\begin{figure}
  \centering
  \includegraphics[width=\linewidth]{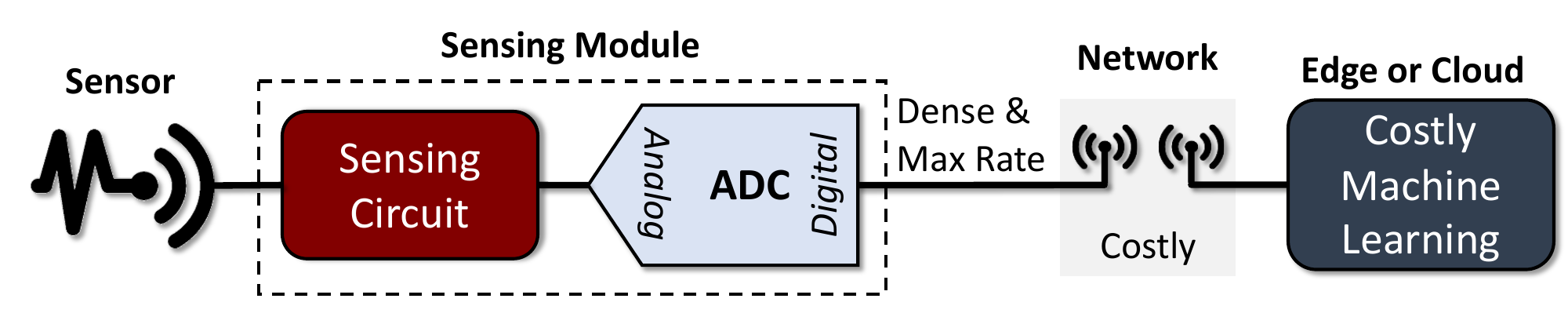}
  \vspace{-8mm}
  \caption{{Existing sensing and information processing pipeline.}}
  \label{Fig:overview_diagram_existing}
  \vspace{-5mm}
\end{figure}

In contrast to today's sensors' dense data generation, biological sensors operate on a significantly smaller scale, generating five orders of magnitude less data through intelligent sensing~\cite{moioli2020neurosciences, mehonic2022brain}. \autoref{Fig:overview_diagram_existing} illustrates the existing sensing systems, where high-precision Analog-to-Digital Converters (ADC) produce highly dense data. Despite proposed intelligent sensing approaches to mitigate massive data generation costs~\cite{hou2023architecting,ma2023camj,ma2023leca}, none have explored controlling high-precision ADC computation activity using low-precision ADC data, known for its energy efficiency~\cite{sarajlic2017low}. Our proposed \emph{Intelligent Sensor Control} introduces a novel concept, leveraging real-time feedback from a machine learning algorithm to ensure sensors generate only necessary data for final analysis, focusing on or producing relevant data for learning purposes.

Challenges in deploying existing deep learning models on or near sensors, including memory consumption and energy-intensive computations, necessitate a paradigm shift. Even lightweight models like YOLOv4 tiny~\cite{bochkovskiy2020yolov4} struggle with radar data~\cite{zhao2023lpdnet}. Our objective is to address these challenges through an intelligent, robust, and efficient framework that represents and analyzes raw sensor data. Moreover, existing learning models struggle to handle raw, noisy low-precision sensor data~\cite{cano2021onlinehd, zou2021scalable}, translating into a non-straightforward integration with sensors. By redesigning machine learning algorithms using neurally-inspired HyperDimensional Computing (HDC)~\cite{Kanerva2009}, we aim to achieve real-time performance with noisy data. HDC, mimicking brain functionalities, offers advantages in efficiency and noise-tolerant computation~\cite{kanerva2009hyperdimensional, zhou2019edge}.

\begin{figure}
  \centering
  \includegraphics[width=\linewidth]{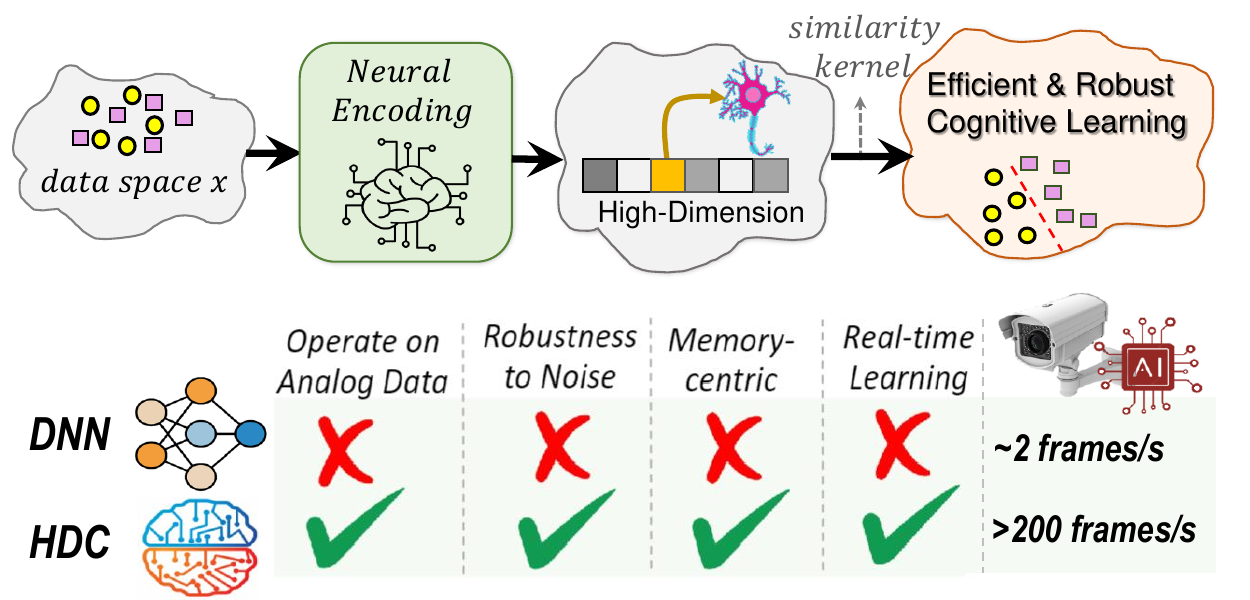}
  \vspace{-8mm}
  \caption{{Characteristics HDC possesses while DNN lacks. These characteristics of HDC make our intelligent sensing more powerful.}}
  \label{Fig:HDDC}
  \vspace{-5mm}
\end{figure}

In this paper, we propose fundamental changes to make sensing systems intelligent for various applications, aiming for four orders of magnitude data reduction through bio-inspired approaches. Our approach draws inspiration from the human visual system, leveraging HDC as an alternative computing method that processes cognitive tasks robustly and efficiently. \autoref{Fig:HDDC} highlights characteristics of HDC that enhance intelligent sensing. Our goal is to render AI more accessible to a wide array of sensing devices by addressing the digital data deluge issue, making sensing systems efficient through HDC's bio-inspired computational model.

Our work is fundamentally novel and provides the following contributions:
\begin{itemize}
    \item To the best of our knowledge, we propose a completely novel concept of Intelligent Sensing, \emph{Intelligent Sensor Control}. Unlike previous works on intelligent sensing that focus on compressing data, our solution selectively generates data, leading to a substantial cost reduction in scenarios where the activity of interest is infrequent.
    \item We propose \emph{HyperSense} model which is capable of conducting HDC-based object detection for enabling visual sensing data based \emph{Intelligent Sensor Control}. We also studied characteristics of \emph{HyperSense} model by thoroughly exploring multiple hyperparameters of the model to identify the optimal parameters suitable for the sensing scenario.
    \item We design an FPGA-based domain-specific accelerator targeting \emph{HyperSense}. To improve the sensing throughput, we customized the computing data path to reuse computations. Our evaluation results show that the FPGA implementation of \emph{HyperSense} achieves on average 5.6$\times$ speedup compared to YOLOv4 running on NVIDIA Jetson Orin, while also delivering improved sensing accuracy with up to 92.1\% energy saving compared to the conventional system.
\end{itemize}


\section{Related Works}

\noindent \textbf{Intelligent Sensing} With the sensor technology improvements over the past decades, computational methods have evolved to help us gather meaningful information from raw sensor data~\cite{ballard2021machine}. Multiple designs from different perspectives have been proposed to improve the data sensing efficiency, including sensor materials~\cite{ballard2021machine}, sensor circuits~\cite{li2021recent}, in-sensor accelerators~\cite{hsu20230,angizi2022pisa,ma2022hogeye}, and near sensor accelerators~\cite{sumbul2022system,zhou2020near}. The key idea of in-sensor and near-sensor acceleration is integrating machine learning computing circuits into the sensing circuit to improve the data processing efficiency. Those supported machine learning kernels include a histogram of gradient (HOG)~\cite{ma2022hogeye}, matrix to matrix multiplication~\cite{hsu20230}, and convolution~\cite{angizi2022pisa}. Although those in-sensor and near-sensor acceleration techniques have significantly improved sensing efficiency, they fail to consider system-level integration. Recently, in computer system and architecture communities, multiple system-level intelligent sensing frameworks have been proposed~\cite{hou2023architecting,ma2023camj,ma2023leca}. Specifically, work in ~\cite{hou2023architecting} focuses on the multi-model computing ($M^2C$) system integration. CAMJ~\cite{ma2023camj} proposes an architecture-level modeling framework of CMOS Image Sensors (CIS). LeCA~\cite{ma2023leca} proposes an in-sensor image compression accelerator to balance the back-end computer vision (CV) model and sensor-capturing image quality. Although, all previous circuit-level and system-level intelligent sensing works mention the importance of ADC in the whole sensing system but none of them try to provide machine-learning solutions to optimize the ADC computation activity. 

\noindent \textbf{Hyperdimensional computing} Brain-inspired hyperdimensional computing (HDC) is based on the understanding that brains compute with patterns of neural activity that are not readily associated with numbers. Due to the huge size of the brain’s circuits, neural patterns can be modeled with hypervectors~\cite{kanerva2009hyperdimensional}. HDC builds upon a well-defined set of operations with random hypervectors, is extremely robust in the presence of failures, and offers a complete computational paradigm that is easily applied to multiple learning problems, such as speech recognition~\cite{imani2017voicehd}, genome sequence alignment~\cite{kim2020geniehd}, graph learning~\cite{kang2022relhd,nunes2022graphhd}, and computer vision~\cite{hersche2022constrained,dutta2022hdnn}. Although HDC-based machine learning models have shown high memorization capability, strong robustness against noise, and nature model interpretability, none of the previous HDC works have tried to solve object detection problems in autonomous driving systems. In this work, we try to use the HDC model to detect objects in radar imaging datasets.  

\section{\emph{HyperSense} Model Design}

\subsection{HDC Basics}~\label{sec:HDC_basic}
The fundamental representational unit of HDC is called a hyperdimensional vector. A hypervector $\mathcal{H}$ indicates a vector $\mathbb{R}^D$ with high dimensionality $D$. The hyperdimensional vectors are compared to each other by a similarity function $\delta$. Utilizing the similarity measure, HDC can facilitate cognitive tasks such as memorization, classification, clustering, and more. HDC frameworks designed to support these tasks rely on three fundamental HDC operations that directly correspond to brain functionalities: bundling, binding, and permutation. Details on each operation are as follows:

\begin{enumerate}
    \item \textbf{Bundling}: this operation, denoted as $+$, is typically implemented as element-wise addition. If $\mathcal{H}=\mathcal{H}_1+\mathcal{H}_2$, then both $\mathcal{H}_1$ and $\mathcal{H}_2$ are similar to $\mathcal{H}$. From a cognitive perspective, it can be interpreted as memorization.
    \item \textbf{Binding}: this operation, denoted as $*$, is typically implemented as element-wise multiplication. If $\mathcal{H}=\mathcal{H}_1*\mathcal{H}_2$, then $\mathcal{H}$ is dissimilar to both $\mathcal{H}_1$ and $\mathcal{H}_2$. Binding also has the important property of similarity preservation in the sense that for some hypervector $\mathcal{V}$, $\delta(\mathcal{V}*\mathcal{H}_1,\mathcal{V}*\mathcal{H}_2)\simeq\delta(\mathcal{H}_1,\mathcal{H}_2)$. From a cognitive perspective, it can be interpreted as association.
    \item \textbf{Permutation}: this operator, denoted as $\rho$, is typically implemented as a rotation of vector elements. Generally, $\delta(\rho(\mathcal{H}),\mathcal{H})\simeq 0$. The permutation is usually used to encode order in sequences.
\end{enumerate}

Using the three basic HDC operations enables a hyperdimensional learning framework for many different tasks. For classification, each step of the framework can be described below.

\begin{enumerate}
    \item \textbf{Encoding}: The first step in the HDC framework is to map the input data $\vec{F}\in U$ into high-dimensional space by introducing an encoding function $\vec{\phi}:U\to H$, which is often referred to as \emph{encoding}. Assume an input vector with $n$ features $\vec{F} = \{f_1, f_2, \dots, f_n\}$ that represents features from voice, image, etc. $\vec{\phi}(\vec{F})=\cos{(\vec{F}\times\vec{B}+\vec{b})}\times\sin{(\vec{F}\times\vec{B})}$, where $\vec{B}$ is a $n\times D$ matrix where every element in $\vec{B}$ is sampled from i.i.d Gaussian distribution ($\mu=0, \sigma=1$), and $\vec{b}$ is sampled from i.i.d uniform distribution over $[0, 2\pi]$. $\vec{\phi}$ preserves some notion of similarity in the input space. Thus, given some input $\vec{x}, \vec{y}\in U$, $\vec{\phi}(\vec{x}),\vec{\phi}(\vec{y})$ are their corresponding hypervector, and $\vec{\phi}(\vec{x})$ is similar to $\vec{\phi}(\vec{y})$ if and only if $\vec{x}$ is similar to $\vec{y}$.
    \item \textbf{Training and Retraining}: Suppose we have a dataset $\mathcal{D}\subset U$ where each data point $\vec{x}_i\in D$ has corresponding label $1 \leq y_i \leq m$ out of $m$ classes. Initial training is done by generating $m$ class hypervectors using bundling: $\vec{C}_i = \sum_{y_j=i}{\vec{\phi}(\vec{x}_j)}$. Based on these class hypervectors, we can start retraining. For each data point to retrain $\vec{x}_i$, each class hypervector is updated as follows:
    $$
    \begin{aligned}
      \vec{C}_l & \leftarrow \vec{C}_l + \eta(1-\delta)\vec{\phi}(\vec{x}_i)\\
      \vec{C}_{l'} & \leftarrow \vec{C}_{l'} - \eta(1-\delta)\vec{\phi}(\vec{x}_i)
    \end{aligned}
    $$
    where $l=y_i, l'\neq y_i$, $\delta=\delta(\vec{C}_l, \vec{\phi}(\vec{x}_i))$, and $\eta$ is learning rate.
    \item \textbf{Inference}: After the class hypervectors $\vec{C}_i$ are updated by the initial training and retraining, given query $\vec{q}\in D$ can be classified in a straightforward way. A class $i$ is considered to be predicted class when $\delta(\vec{C}_i, \vec{\phi}(\vec{q})) > \delta(\vec{C}_j, \vec{\phi}(\vec{q}))$ is satisfied for all $j\neq i$.
\end{enumerate}

\subsection{HDC Intelligent Sensor Control Framework}

\begin{figure}
  \centering
  \includegraphics[width=\linewidth]{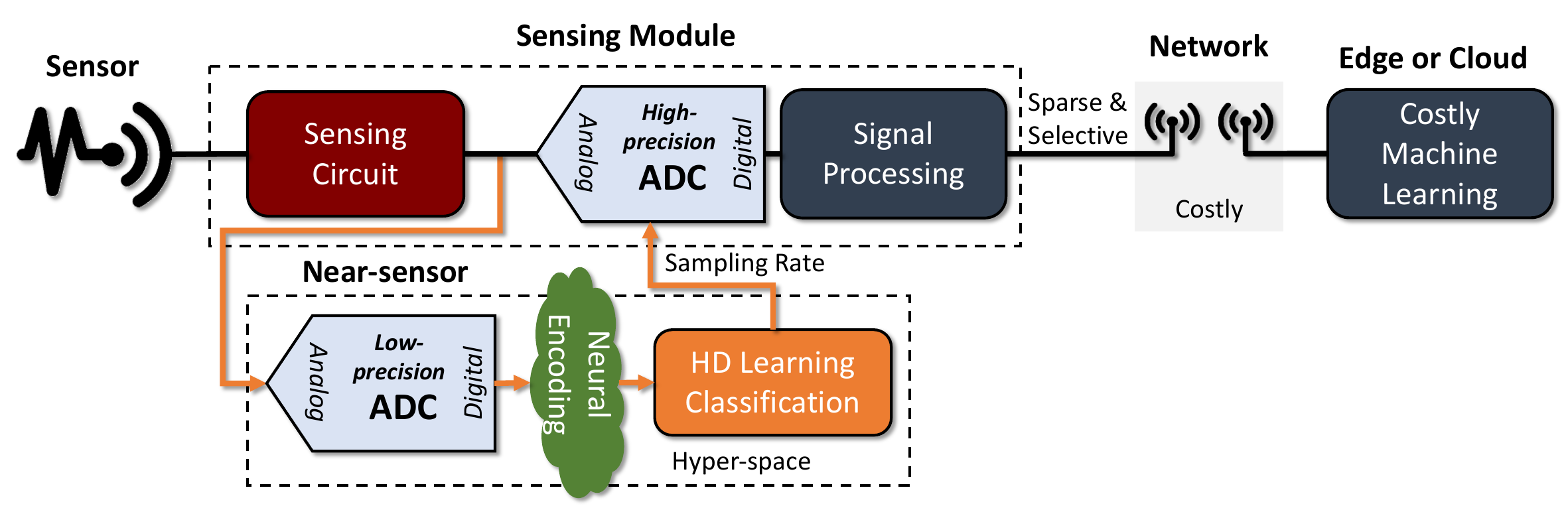}
  \vspace{-8mm}
  \caption{{Overview of our Intelligent Sensing pipeline.}}
  \label{Fig:overview_diagram_ours}
  \vspace{-3mm}
\end{figure}

\begin{figure}
  \centering
  \includegraphics[width=\linewidth]{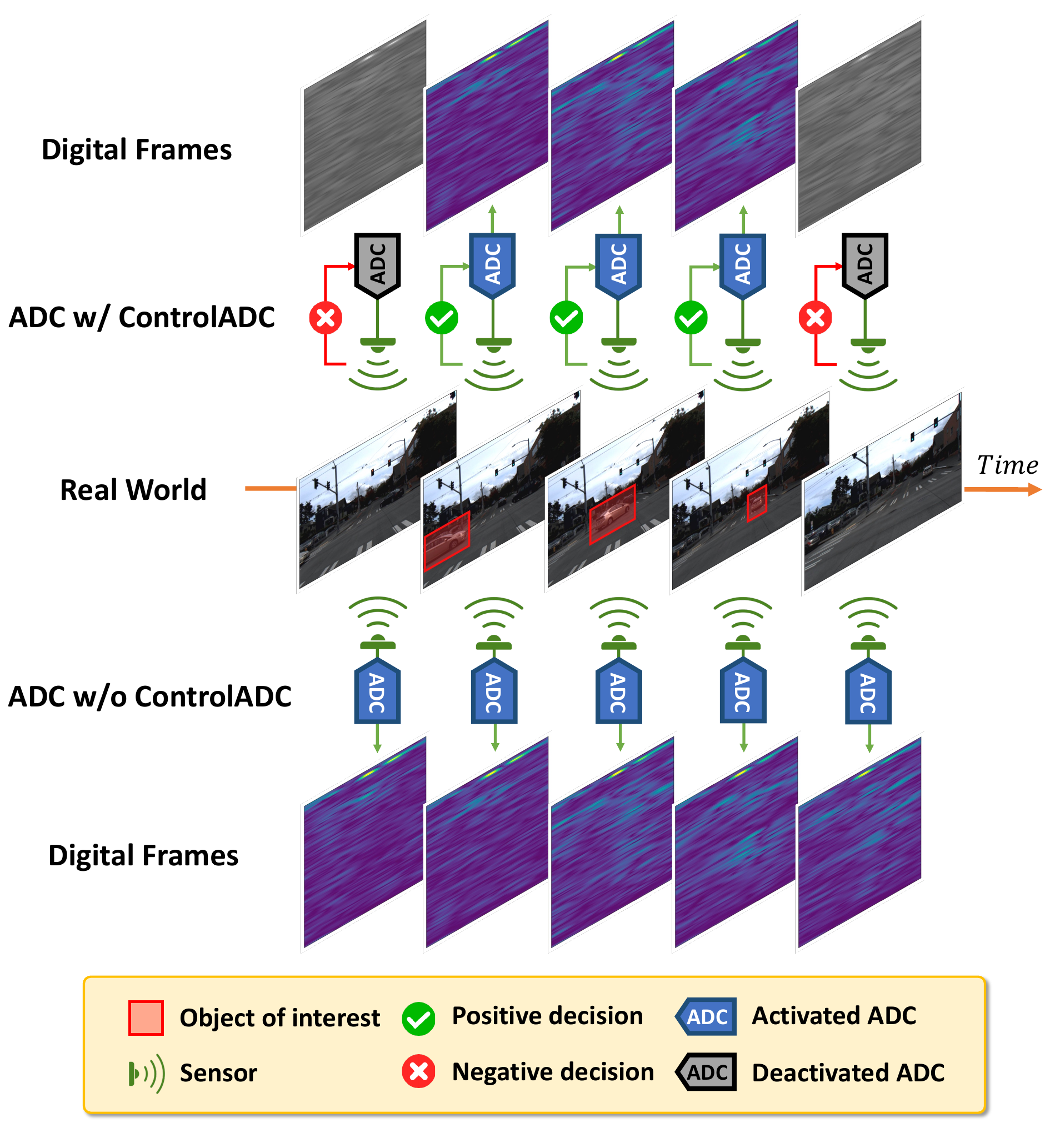}
  \vspace{-10mm}
  \caption{{Illustration of how our proposed framework disables ADC to prevent the excessive generation of digital frames. Generating digital frames with no useful information cause an unnecessary increased cost in the system without \emph{HyperSense}.}}
  \label{Fig:HyperSense_framework_demo}
  \vspace{-5mm}
\end{figure}

\autoref{Fig:overview_diagram_ours} presents an overview of our framework, which leverages HDC to enable intelligent sensing by tightly integrating it with the sensing circuit. The ADC module, responsible for converting analog data into the digital domain, is the most power-consuming and latency-inducing component of every sensor. To achieve efficient and intelligent sensing, our HDC algorithms operate over low-precision digital data generated from energy-efficient low-precision ADC and provide real-time feedback for selective sampling, reducing the data generation rate from the sensor.

Our proposed framework comprises two major components: (1) A neural encoding module that receives raw low-precision data and transforms it into holographic vectors in a high-dimensional space. These hypervectors store information in their patterns and are learnable by our HDC algorithms. (2) A learning algorithm that makes decisions about the sampling rate of the ADC block. The HDC learning aims to lower the ADC sampling rate for data points that do not carry useful information. For instance, our sensing circuit, which typically generates 60 samples/second, would only generate data at a minimum frequency (e.g., 1 frame/second) unless HDC detects that the incoming data points carry relevant information.

Our framework primarily focuses on visual sensing data, particularly radar data, where useful information exhibits locality. As such, the HDC learning and classification algorithm in our framework performs object detection tasks to determine the presence of any object in a given frame of sensing data. \autoref{Fig:HyperSense_framework_demo} illustrates how our framework operates in controlling the ADC. In the figure, an object of interest appears in the second frame and disappears in the last frame. Since the first and last frames do not contain any objects of interest, our framework, shown at the top of the figure, disables the ADC from generating digital frames. However, at the bottom of the figure, without our \emph{HyperSense} model, digital frames are generated even without any objects of interest, resulting in an abundance of useless data that increases the overall system's cost.

\subsection{HDC Object Detection}~\label{sec:HD_object}

\begin{figure*}[t!]
  \centering
  \includegraphics[width=\linewidth]{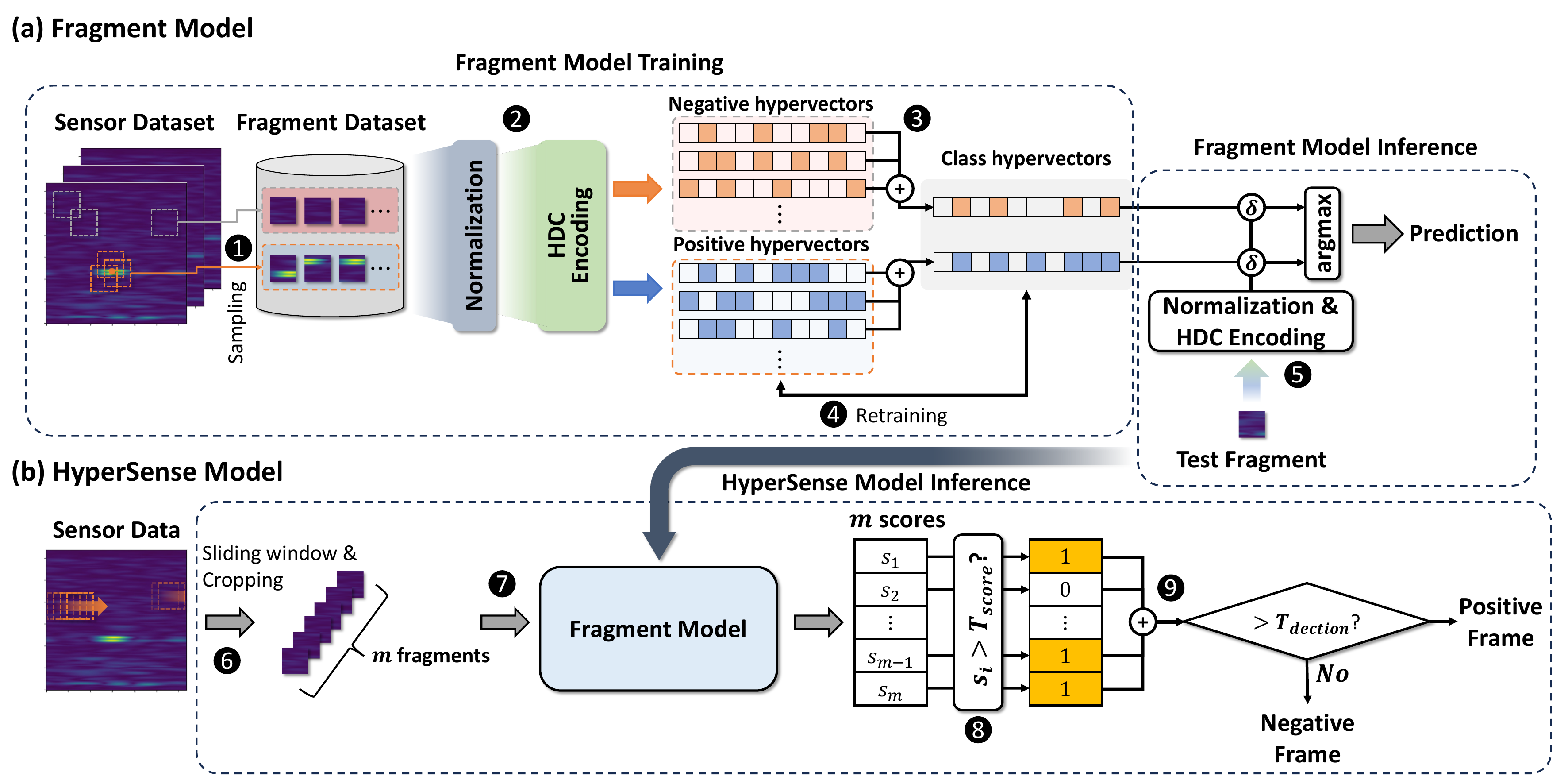}
  \vspace{-8mm}
  \caption{{Overview of our object detection framework for Intelligent Sensing. The object detection framework consists of two models: (a) \emph{Fragment model} and (b) \emph{HyperSense model}. The trained \emph{Fragment model} is applied to the \emph{HyperSense model}.}}
  \label{Fig:framework_pipeline}
  \vspace{-5mm}
\end{figure*}

\begin{figure} 
    \centering
    \includegraphics[width=1\linewidth]{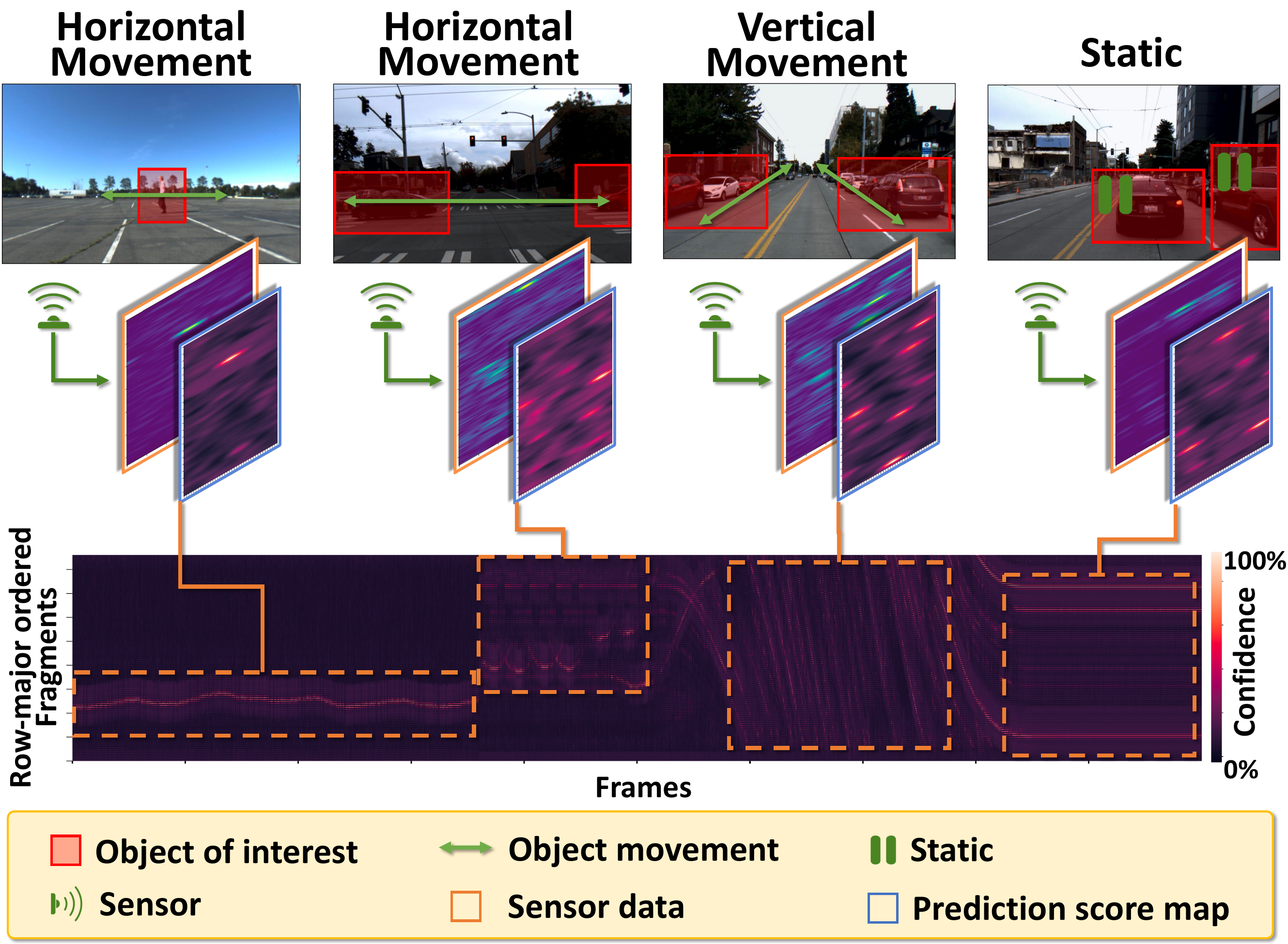}
    \vspace{-7mm}
    \caption{Demonstration of how proposed \emph{HyperSense} model works with \emph{Fragment model} in different types of scenes of a real dataset. Each type of scene presented the top of the figure has a corresponding region in the bottom heatmap presenting prediction scores of \emph{Fragment model} for fragments generated by \emph{HyperSense} model. Each column in the heatmap contains scores for all fragments at a single frame. Scores are ordered in a row-major fashion.}
    \label{fig:object_detection_demo}
    \vspace{-7mm}
\end{figure}

As we mentioned above, our framework controls the frequency of ADC by detecting objects in a given sensing data with locality features. \autoref{Fig:framework_pipeline} shows an overview of our object detection framework. It consists of two models: \emph{Fragment model} and \emph{HyperSense model}. \emph{Fragment model} is designed to perform prediction over a single fragment using hyperdimensional classification. \emph{HyperSense model} uses trained \emph{Fragment model} to conduct prediction over a single frame of sensor data.

Since \emph{HyperSense model} uses trained \emph{Fragment model}, it starts with training \emph{Fragment model} as shown in the \autoref{Fig:framework_pipeline}.(a). First, given a sensor dataset $D_s$, we generate a fragment dataset $D_f$ from the sensor dataset by random sampling positive and negative fragments from each frame of sensor data in a way that positive samples $f_{pos}\subset D_f$ contain object positions while negative samples $f_{neg}\subset D_f$ do not (\myCircled{1}). It is also important to balance the number of negative and positive samples. After we have the dataset $D_f$ to train an HDC classification model, we normalize and encode each data $x_i\in D_f$ (\myCircled{2}). To normalize 2-dimensional fragment $x_i$, $x_i$ is flattened into 1-dimensional vector $\vec{x}_i$ by concatenating each row horizontally. Then, normalized fragment vector $\vec{x}_i'$ is computed by $\vec{x}_i'=\frac{\vec{x}_i}{\|\vec{x}_i\|_2}$. HDC encoding is conducted on $\vec{x}_i'$ to have hypervector $\vec{H}_i = \phi(\vec{x}_i')$ that corresponds to the hyper-dimensional data point of the fragment $x_i$. The 5K or 10K dimensionality is usually selected, however, it highly varies depending on the input data, thus needs to be tuned to have optimized dimensionality balancing between performance and efficiency. By performing the normalization and HDC encoding process for all negative and positive fragments, we can have negative hypervectors and positive hypervectors respectively. Now, initial training of HDC classification proceeds to have class hypervector by bundling each negative $\vec{C}_{neg}=\sum_{x_i\in f_{neg}}{\vec{H}_i}$ and positive hypervectors $\vec{C}_{pos}=\sum_{x_i\in f_{pos}}{\vec{H}_i}$ (\myCircled{3}). To have a better-performing HDC classification model, retraining using the negative and positive hypervectors is conducted in an iterative way (\myCircled{4}). The best-performing HDC model is selected by running inference over test fragments to compute performance metrics such as accuracy, f1 score, etc. (\myCircled{5}). For each test fragment $q$, the inference is conducted by computing the similarity between each class hypervector with normalized and HDC encoded hypervector of $q$, which can be formalized by $\delta(\vec{C}_i, \phi(\vec{q}'))$. And the class with the higher similarity value is considered to be predicted. With the best performing HDC model, we now have trained \emph{Fragment model}.

Using the trained \emph{Fragment model}, we can have \emph{HyperSense model} as shown in the \autoref{Fig:framework_pipeline}.(b). In order to form a \emph{HyperSense model}, not only the trained \emph{Fragment model}, we also need to set additional hyperparameters $T_{score}$, $T_{detection}$, and $stride$. As long as we have the four of them, we do not need additional training steps for \emph{HyperSense model}. We will discuss each of the hyperparameters in the following. In a given frame of sensor data, inference over the sensor frame using \emph{HyperSense model} is starting with cropping fragments in a sliding window manner (\myCircled{6}). In this step, the hyperparameter $stride$ is used. The sliding window moves over the frame by $stride$ amount at a time for both horizontal and vertical movement. Then, we give the resulting $m$ fragments $f_i$ to the trained \emph{Fragment model} to have prediction scores $s_i$ for each fragment $f_i$ (\myCircled{7}). Now, we will have $m$ scores each corresponding to a fragment in a given frame of sensor data. In order to determine which fragments are detected objects, the hyperparameter $T_{score}$ is used (\myCircled{8}). Each score $s_i$ is compared with the score threshold $T_{score}$ to determine the prediction of objects in a fragment $f_i$. If it is larger than $T_{score}$, it is considered to have objects in the fragment, and store prediction value of $1$ otherwise $0$. Finally, we compare the summation of all prediction values with the hyperparameter $T_{detection}$ which indicates the threshold value for the number of detection (\myCircled{9}). If the summation value is larger than $T_{detection}$, the final prediction of \emph{HyperSense model} is positive which indicates there are objects in a given frame of sensor data otherwise, the prediction is negative which indicates there is no objects in a given frame of sensor data.

\begin{figure*}[t!]
  \centering
  \includegraphics[width=1\linewidth]{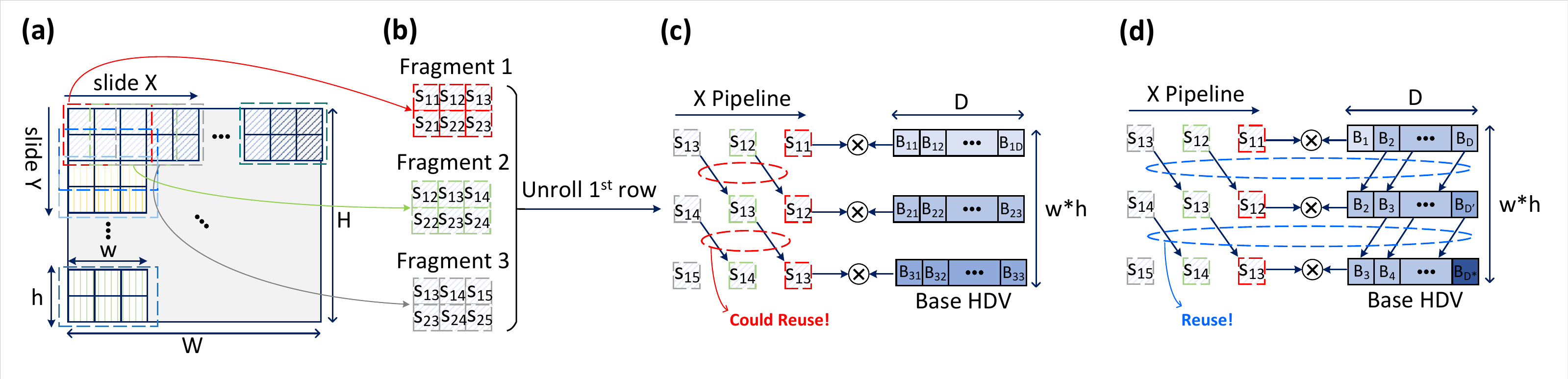}
  \caption{{(a) Input image Fragmentation. (b) Fragment Encoding Pipeline. (c) The HDC encoding process of fragment. (c) The encoding computation reuse.}}
  \label{Fig:motivation}
\end{figure*}

\autoref{fig:object_detection_demo} illustrates an actual demonstration conducted on the CRUW dataset, which will be further explained in a later section, showcasing how our \emph{HyperSense} model interacts with the trained \emph{Fragment model}. The demonstration consists of four different types of scenes, each with a corresponding area in the heatmap displayed at the bottom of the figure. The heatmap represents the prediction scores of the \emph{Fragment model}, where each fragment corresponds to the y-axis, and each time frame corresponds to the x-axis. The fragments are organized in a single column through row-major ordering, with the score for the topmost left fragment located at the top of each column in the heatmap.
In scenes involving horizontal movements, we observe high confidence values arranged horizontally with little vertical movements, while in scenes with vertical movements, they are vertically located. Moreover, in the static scene, we can observe consistent scores across the time frames.

\section{Accelerator Architecture Design}

\subsection{Computation Bottleneck}

\begin{figure*}
    \centering
    \includegraphics[width=1\linewidth]{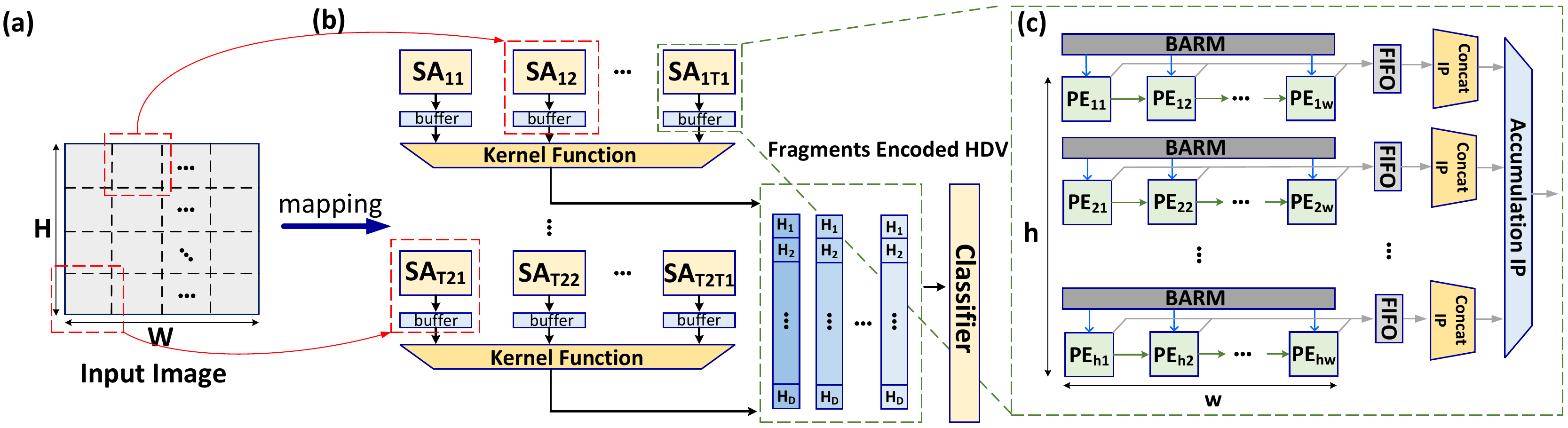}
    \vspace{-7mm}
    \caption{The top-level architecture design. (a) Input radar image partition. (b) Systolic array (SA) groups. (c) Processing elements (PE) interconnect.}
    \label{fig:architecture}
    \vspace{-5mm}
\end{figure*}

Though HDC-based models show promising potential in conducting object detection tasks, to perform real-time learning, it is still necessary to design domain-specific accelerators to avoid unnecessary computations. Specifically, we illustrate the computation overhead in \autoref{Fig:motivation}. As is discussed in \autoref{sec:HD_object}, to detect an object, the HDC model needs to encode each fragment by mapping input image data from normal space into hyperspace. As is shown in \autoref{Fig:motivation}.(a), the sliding window will scan the input image in both the X direction and the Y direction. For simplicity, here we only zoom into the situation of X direction. To maximize the computation efficiency, the most straightforward optimization strategy is to pipeline the encoding operations of different fragments in one direction as shown in \autoref{Fig:motivation}.(b). Here we suppose the dimension of the sliding window is $2\times3$ and the step size of the scanning operation is 1. To encode the fragment, we need to first unroll the fragments from $2\times3$ into $6\times1$ and then multiply each element of the unrolled vector with base hypervector as we discussed in \autoref{sec:HDC_basic}. Here in~\autoref{Fig:motivation}.(c), we present the encoding process of the first rows of fragment $S_1$, $S_2$, and $S_3$. However, as we show in the \autoref{Fig:motivation}.(c), fragments share the same elements with each other. For example, fragment $S_1$ and fragment $S_2$ will share 2 elements in the first row. Therefore, to improve the computation efficiency, one of the optimization strategies is to reuse the computation of common elements between different fragments. 

\subsection{HDC Encoding Optimization}\label{sec:encoding_resue}
To reuse the common elements between continuous fragments, here we use the attributes of permutation of different base hypervectors. As is discussed in~\autoref{sec:HDC_basic}, to keep the orthogonality of different base hypervectors, one of the ways is first randomly generating the first base hypervector and then using HDC permutation operation to generate the rest base hypervectors. Suppose the size of the fragment is $h \times w$, and the base hypervector matrix is \textbf{B}. If the hypervector dimension is D then the size of the base hypervector matrix \textbf{B} is $h \times w \times D$. To achieve the balance between efficiency and accuracy, we randomly generate the hypervector of the first axis of the matrix \textbf{B} and use HDC permutation operation to generate the rest in the second axis. In mathematics, this process can be represented as:
\begin{equation}
    \Vec{B}_{i,j+1} = perm(\vec{B}_{i,j}) \quad where \; i\in [1,h] \; and \; j\in [1,w-1]
\end{equation}

An example of a permutation base hypervector generation process is shown in \autoref{Fig:motivation}.(d). Here we suppose the base hypervector of element $S_{12}$ is left shifted by the base hypervector of element $S_{11}$. The same principle can be also applied to the generation of 
the base hypervector of element $S_{13}$ which is left shifted by the base hypervector of element $S_{12}$.

Here we use a simple example to illustrate how to reuse the computation in~\autoref{Fig:motivation}.(d). It is easy to notice that the second element of $S_1$ is the same as the first element of $S_2$. The HDC encoding computation of the first-row fragment $S_1$ could be represented as:
\begin{equation}
    \vec{H_1} = S_1\times \vec{B_1} + S_2\times \vec{B_2} + S_3\times \vec{B_3}
\end{equation}
The encoding computation of the first row of fragment $S_2$ could be represented as:
\begin{equation}
    \vec{H_2} = S_2\times \vec{B_2} + S_2\times \vec{B_3} + S_3\times \vec{B_4}
\end{equation}
Suppose the base hypervector $\vec{B_2}$ is left shifted by $\vec{B_1}$ and base hypervector $\vec{B_3}$ is left shifted by $\vec{B_2}$. In this case, we can rewrite HDC encoding computation as:
\begin{equation}
    \vec{H_1} = S_1\times[B_1,B_2,...] + S_2\times[B_2,B_3,...] + S_3\times[B_3,B_4,...]
\end{equation}
\begin{equation}
    \vec{H_2} = S_2\times[B_1,B_2,...] + S_3\times[B_2,B_3,...] + S_4\times[B_3,B_4,...]
\end{equation}
As is shown in~\autoref{Fig:motivation}.(d), for those overlapping elements of each fragment, the D-1 multiplications result could be reused to save the computation. Therefore, the overall encoding overhead is significantly reduced. Here we want to mention that the computation reuse is based on the fact that HDC base hypervectors will still keep holography property when applying permutation operations to generate base hypervector. This attribute is unique in the HDC model and cannot be used for traditional neural network-based models.

\subsection{Accelerator Architecture Design}\label{sec:architecture}

In~\autoref{fig:architecture}, we present the architecture design of HDC near the sensor processing accelerator. For each input image frame, we first partition it into several small pieces, as is shown in~\autoref{fig:architecture}.(a). Each small piece will be mapped into a systolic array (SA) IP for HDC encoding operations and be processed parallel as is shown in~\autoref{fig:architecture}.(b). Here we suppose the original input image's dimension is $H \times W$. There are total $T_1 \times T_2$ SA IPs inside the near sensor accelerator. Inside each SA IP, the sliding window will move in both X and Y directions to generate multiple fragments. Each fragment inside a single SA IP will be generated in pipeline style and different SA IPs will run parallel. In the end, an HDC classifier will perform a cosine similarity check to determine whether the object is detected. For FPGA on-chip kernel function and classifier implementation, we refer to previous HDC FPGA work~\cite{imani2021revisiting}.

\begin{figure}
    \centering
    \includegraphics[width=1\linewidth]{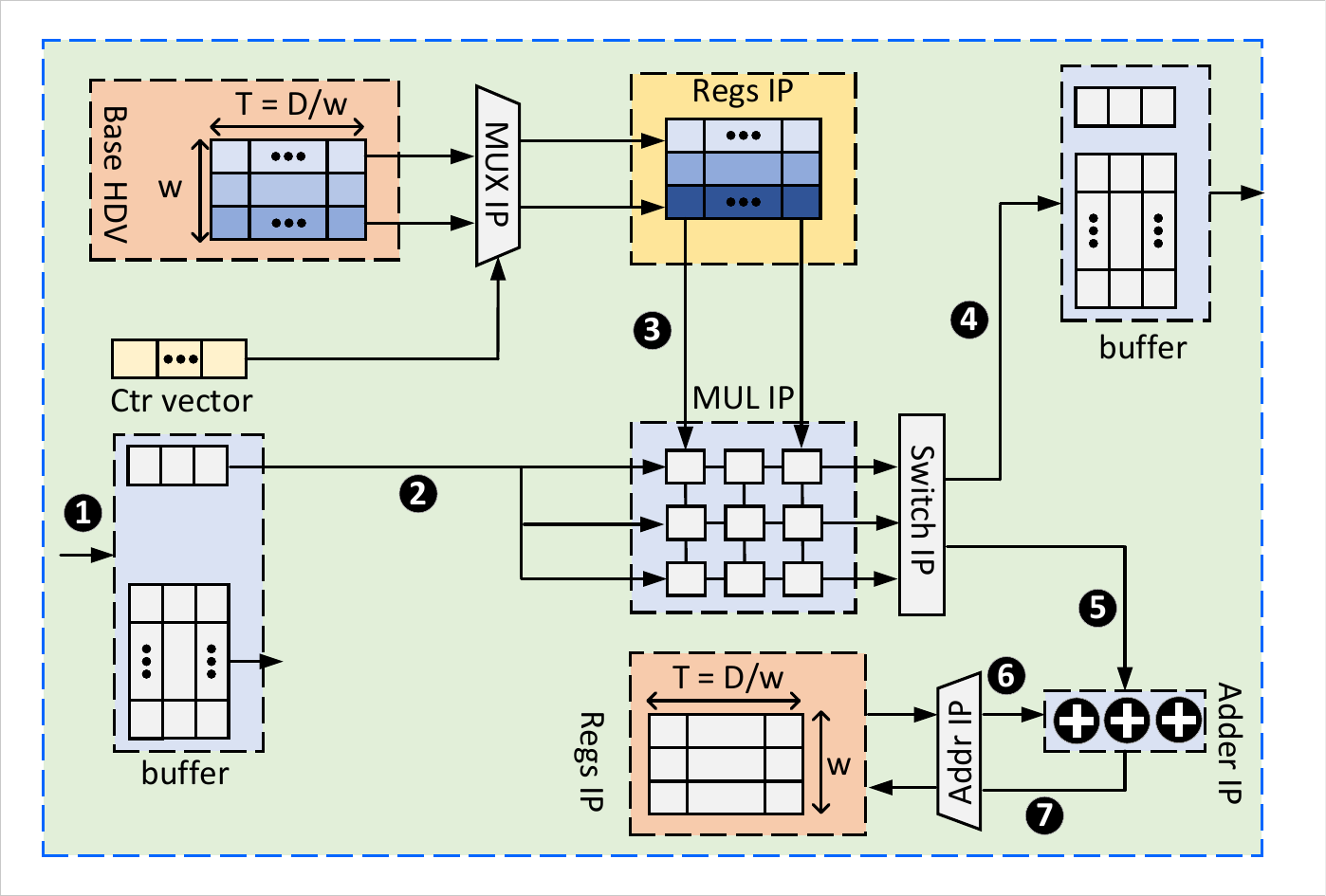}
    \vspace{-7mm}
    \caption{Microarchitecture design of process elements IP.}
    \label{fig:PE_microarchitecture}
    \vspace{-5mm}
\end{figure}

In~\autoref{fig:architecture}.(c), we present the architecture design of each SA IP. Suppose the dimension of each fragment is $h\times w$, in this case, each SA IP will consist of $h\times w$ process element (PE) IPs. Those PE IPs in the same row will execute in pipeline style and PE IPs in the different rows will run parallel. Suppose the hypervector dimension is D, then each PE IP will perform encoding operations of hypervector chunks whose dimension is $\frac{D}{w}$. In~\autoref{fig:PE_microarchitecture}, we present each PE IP's microarchitecture. To realize computation reuse, each PE IP will accept the image element and partial encoding result from the left PE IP and buffer it inside its FIFO IP, as is shown in~\autoref{fig:PE_microarchitecture} \myCircled{1}. The input image element will multiply with base hypervector chunks (\myCircled{2} and \myCircled{3}). As we will see in ~\autoref{sec:reuse_example}, only part of the base hypervector chunks will participate in the multiply operations. After the multiplication, a partial of the encoding result will pass into the next PE IP (\myCircled{4}). All the new encoding results of the current PE IP (\myCircled{5}) together with partial encoding results from the last PE IP (\myCircled{6}) will be used to update encoding hypervector chunks at registers IP (\myCircled{7}).

\begin{figure*}
    \centering
    \includegraphics[width=1\linewidth]{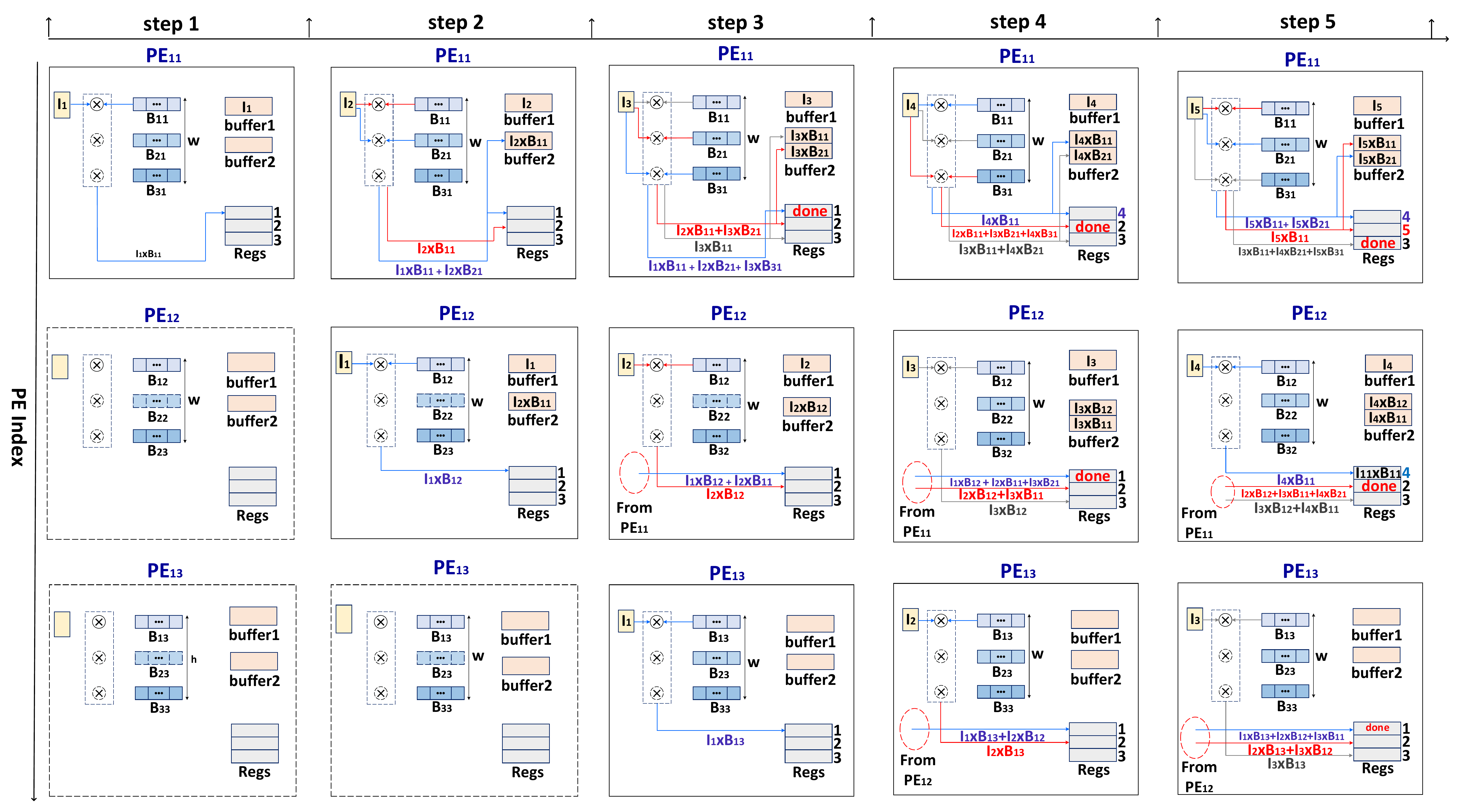}
    \vspace{-7mm}
    \caption{HDC fragments encoding computation reuse example. Here we suppose the width of the sliding window is 3. Due to page size limitation, only 5 steps are presented in this figure. }
    \label{fig:dataflow_example}
    \vspace{-7mm}
\end{figure*}

\subsection{Computation reuse example} \label{sec:reuse_example}
In~\autoref{fig:dataflow_example}, we present an example of HDC encoding computation reuse. Due to page restriction, here we limit the dimension of the fragment as $2\times3$. As is discussed in~\autoref{sec:architecture}, we parallel the computation of each row of the fragment and reuse the computation of the elements in a single row. Therefore, in~\autoref{fig:dataflow_example}, we show the computation activity of the first row of SA IP which includes three PE IPs. 

Suppose the hyperspace dimension is D, then inside each PE IP, the chunk vector dimension is $\frac{D}{3}$. We use $P_{ij}$ to represent the $i^{th}$ row and $j^{th}$ column PE IP inside single SA IP. In~\autoref{fig:dataflow_example}, as we only present the first row of SA IP's computing activity, all PE IP's i index is 1. Due to page limitation, we only show 5 stages of the pipeline process but for longer time steps, the principle is still the same. Here we use $\Vec{B_{ij}}$ to represent the base encoding hypervector. Specifically, i represent the position and j represents the chunk index. Therefore, for the first row of the $k^{th}$ fragment, the encoding computation mathematics representation of the $m^{th}$ hypervector chunk should be:

\begin{equation}
    \vec{H_{km}} = I_{k}\times\Vec{B_{1m}} + I_{k+1}\times\Vec{B_{1m}} + I_{k+2}\times\Vec{B_{1m}} \quad m\in[1:3]
\end{equation}

In ~\autoref{fig:dataflow_example}, we show the computation reuse of fragment 1, fragment 2, and fragment 3 whose computation mathematics of the $m^{th}$ chunk should be:
\begin{equation}
    \vec{H_{1m}} = I_{1}\times\Vec{B_{1m}} + I_{2}\times\Vec{B_{1m}} + I_{3}\times\Vec{B_{1m}} \quad m\in[1:3]
\end{equation}
\begin{equation}
    \vec{H_{2m}} = I_{2}\times\Vec{B_{1m}} + I_{3}\times\Vec{B_{1m}} + I_{4}\times\Vec{B_{1m}} \quad m\in[1:3]
\end{equation}
\begin{equation}
    \vec{H_{3m}} = I_{3}\times\Vec{B_{1m}} + I_{4}\times\Vec{B_{1m}} + I_{5}\times\Vec{B_{1m}} \quad m\in[1:3]
\end{equation}

As we discussed in~\autoref{sec:encoding_resue}, to reuse the encoding computation of different hypevector chunks, we generate the base hypervector based on HDC permutation operation. Specifically, for the first base hypervector $\vec{B_1}$, all three chunks $\vec{B_{11}}$, $\vec{B_{12}}$ and $\vec{B_{13}}$ are generated based on Gaussian distribution as is discussed in~\autoref{sec:HDC_basic}. For base hypervector $\Vec{B_2}$, we have:
\begin{equation} \label{eq:base_permutation}
    \Vec{B_{22}} = \Vec{B_{11}} \quad and \quad \Vec{B_{23}} = \Vec{B_{12}}
\end{equation}
This means the second base hypervector $\Vec{B_2}$ is generated by applying permutation operation for the first base hypervector $\Vec{B_1}$. The first chunk of $\Vec{B_2}$ ($\Vec{B_{21}}$) is still generated with Gaussian distribution. For the base hypervector $\Vec{B_3}$, we have:
\begin{equation}
    \Vec{B_{32}} = \Vec{B_{21}} \quad and \quad \Vec{B_{33}} = \Vec{B_{22}}
\end{equation}
Next, we will discuss the computation reuse step by step.
As is shown in~\autoref{fig:dataflow_example} step 1, the first element of the fragment $S_1$ ($I_1$) will be first multiplied with the first chunk of the first base hypervector ($B_{11}$) inside $PE_{11}$. After the multiplication operation, the temporal encoding result will be saved inside each PE IP's registers (Regs). Also, $I_1$ will be buffered inside $PE_{11}$'s buffer which is connected with $PE_{12}$. At the first stage of the pipeline, only $PE_{11}$ is active and the other two PE IPs are idle. 

At step 2, element $I_2$ will be loaded into $PE_{11}$ IP and element $I_1$ will be passed into $PE_{12}$ IP. Since $I_2$ simultaneously correspond to the first element of fragment $S_2$ and the second element of fragment $S_1$, inside $PE_{11}$, $I_2$ will multiply with both $B_{11}$ and $B_{21}$. The multiplication result of $I_2 \times B_{11}$ will be saved inside the second row of Regs and the result of $I_2 \times B_{21}$ will update the first row of Regs. After update the temporal fragment $S_1$ encoding result inside $PE_{11}$ will be $I_1 \times B_{11} + I_2 \times B_{21}$. To reuse the computation in stage 3, we also buffer the result of $I_2 \times B_{11}$. We will reuse the computation result of $I_2 \times B_{11}$ at stage 3. During the computing activity of $PE_{11}$, there is also another computation inside $PE_{12}$, where element $I_1$ is multiplied with base hypervector chunk $B_{12}$. $I_1 \times B_{12}$ will also be saved inside the first row of Regs as the partial encoding result of fragment $S_1$. During step 2, the $PE_{13}$ is still idle. 

At step 3, element $I_3$ will be loaded into $PE_{11}$, element $I_2$ will be forward into $PE_{12}$, and element $I_{13}$ will be passed into $PE_{13}$ IP. Starting from element $I_3$, all elements will be shared by three fragments, which means it needs to be multiplied with all 3 base hypervectors. So at step 3, inside $PE_{11}$, we compute the result of $I_3 \times B_{11}$, $I_3 \times B_{21}$, and $I_3 \times B_{31}$. Like step 2, $I_3 \times B_{11}$ is a partial result of fragment $S_3$ which will be saved inside the third rows of Regs. $I_3 \times B_{31}$ and $I_3 \times B_{21}$ will be used to update partial result of fragment $S_1$ and fragment $S_2$ respectively. Until now, we have finished the first chunk encoding operation of fragment $S_1$ which is $\Vec{H_{11}}$. We will pop out $\vec{H_{11}}$, buffer it and waiting for the result of $\Vec{H_{12}}$ and $\Vec{H_{13}}$. The $PE_{12}$ IP will simultaneously accept element $I_2$ and partial encoding result $I_2 \times B_{11}$. Based on~\autoref{eq:base_permutation}, we have:
\begin{equation}
    I_2 \times \Vec{B_{11}} = I_2 \times \Vec{B_{22}} \quad since \quad \Vec{B_{11}} = \Vec{B_{22}}
\end{equation}
So we directly use $I_2 \times B_{11}$ coming from $PE_{11}$ to update the partial encoding result of $S_1$ and only perform the calculation of $I_2 \times B_{12}$ which will be used as a partial result as fragment chunk encoding result of fragment $S_2$. Since $ \vec{B_{12}} = \Vec{B_{23}}$, to reuse the computation in stage 4, we also buffer the result of $I_2 \times B_{12}$ which will be passed into $PE_{13}$. 

After step 3, the later stages' computation activities are the same. At the $PE_{11}$ IP, the input element needs to be multiplied with all base hypervector chunks. To reuse the computation result, the input element's multiplication result with the first two base hypervector chunks ($\Vec{B_{11}}$ and $\Vec{B_{21}}$) will be buffered and loaded into the later PE IPs at the next stages. For other IPs ($PE_{12}$ and $PE_{13}$), the input element only needs to be multiplied with the first base hypervector chunk ($\Vec{B_{21}}$ and $\Vec{B_{31}}$), Meanwhile, the computation of those multiplications gonna be reused by next PE IPs. As is shown in~\autoref{fig:dataflow_example}, as we adopt the HDC permutation operations to base hypervectors, the computation is significantly reduced.

\section{Experiments}

\subsection{Experimental Setup}

The proposed framework has been executed with a software framework and a hardware accelerator. 
Our software framework is implemented using a combination of Pytorch and NumPy that supports HDC encoding and classification. We study the effectiveness of our technique over the CRUW dataset~\cite{wang2021rethinking}, which is a public camera-radar dataset for autonomous vehicle applications. The radar images are captured by TI AWR1843 whose operating power is around 30W~\cite{wang2021rethinking,li2022real}. For consistency and simplicity, we limited our experiments to considering square shapes for the fragments. Thus, when we refer to a fragment size of $x$, it corresponds to $x\times x$ sized square fragments.  For the hardware accelerator, we implemented our design using SystemVerilog and tested it on Xilinx Zynq UltraScale + MPSoC ZCU104 (Xilinx). Our accelerator architecture design is platform agnostic and can be implemented on both FPGA and ASIC. In this paper, we choose to use FPGA as the evaluation platform to quickly test our design's efficiency. We leave ASIC evaluation as our future work.

\subsection{Evaluation of \emph{HyperSense} model}

First, we focused on a scenario where we aim to allow a specific amount of false positives while maximizing the true positive rate (TPR). To analyze this scenario and determine the setting of our models that achieves the maximum TPR at the desired false positive rate (FPR), we employed the Receiver Operating Characteristic (ROC) curve evaluation. By plotting the ROC curves for different model configurations, we can identify the optimal operating points that strike the right balance between true positives and false positives, enabling us to achieve the highest TPR while adhering to the desired FPR. This analysis provides valuable insights into the performance and effectiveness of our models under different settings, ensuring their efficiency and suitability for intelligent sensing applications.

\begin{figure}
    \centering    \includegraphics[width=1\linewidth]{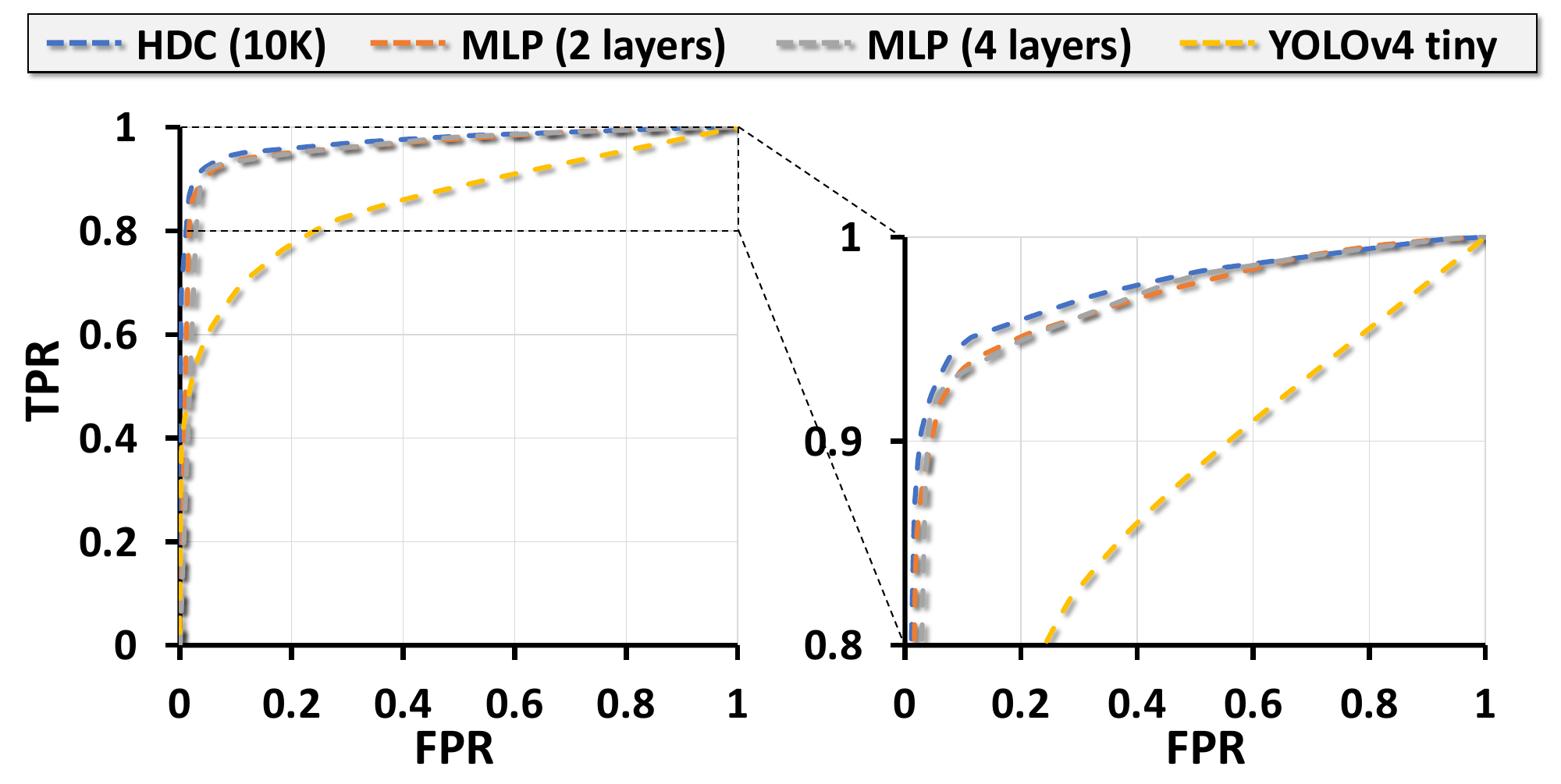}
    \vspace{-8mm}
    \caption{Testing \emph{Fragment model} with other baseline models that widely used lightweight models for object detection when fragment size is 128.}
    \label{fig:fragment_model_compare}
    \vspace{-4mm}
\end{figure}

\begin{table}[]
\caption{Area Under the Curve (AUC) comparison with different baseline models including our \emph{Fragment model} when considering True Positive Rate (TPR) larger than 0.8.}
\label{tab:auc_compare}
\vspace{-3mm}
\resizebox{0.48\textwidth}{!}{%
\begin{tabular}{c|cccc}
\toprule
              & HDC (10K)    & MLP (2 layers)                        & MLP (4 layers)   & YOLOv4 tiny    \\ \midrule
        AUC      & \textbf{0.1739}    & 0.1685                        & 0.1681   & 0.0803
              \\ \bottomrule
\end{tabular}%
}
\vspace{-7mm}
\end{table}

\begin{figure}
    \centering
    \includegraphics[width=1\linewidth]{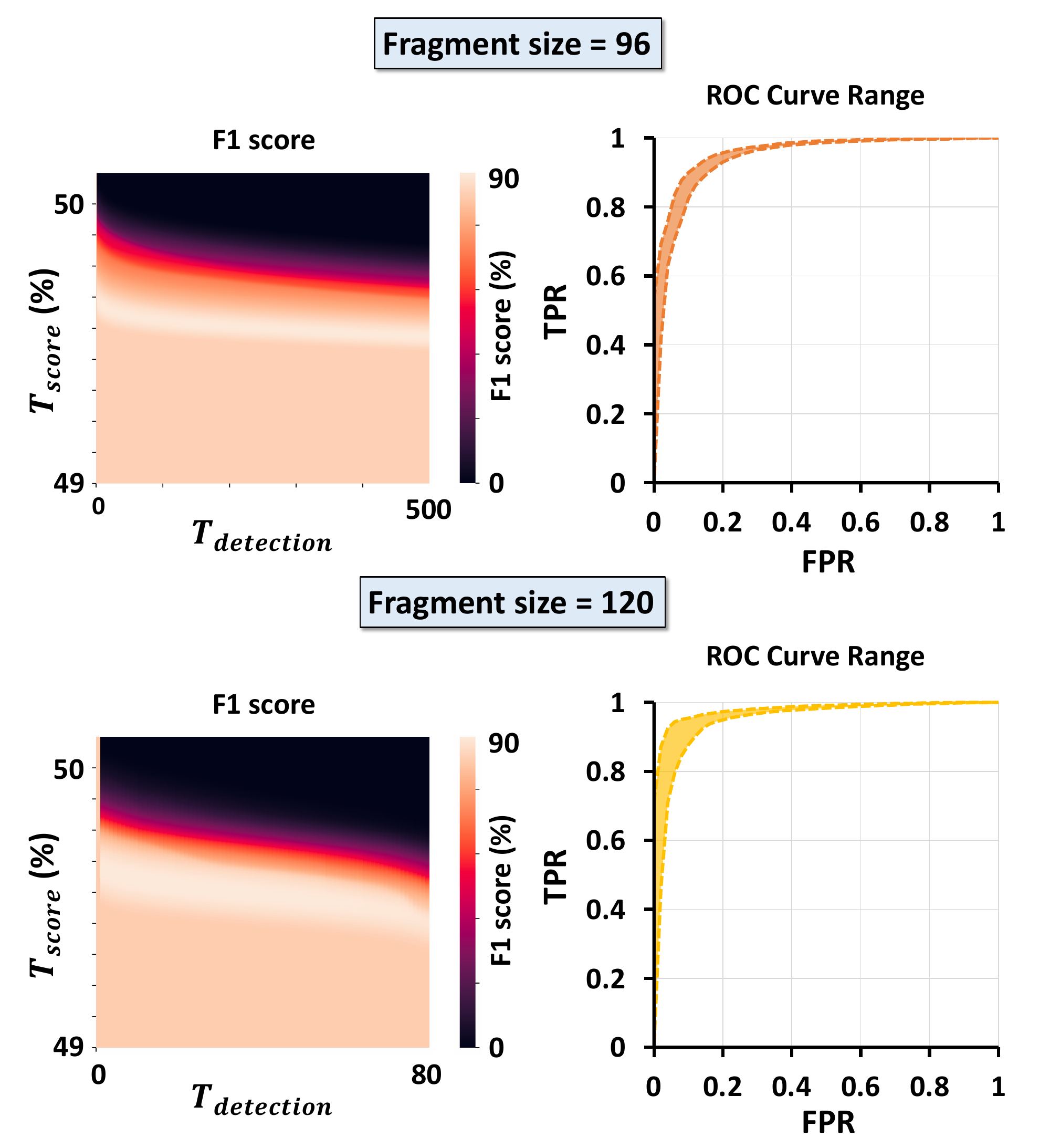}
    \vspace{-9mm}
    \caption{Exploration of two thresholds of \emph{HyperSense} model $T_{score}$ and $T_{detection}$ with their effect on ROC Curve.}
    \label{fig:thresholds_test}
    \vspace{-3mm}
\end{figure}

\begin{figure}
    \centering
    \includegraphics[width=1\linewidth]{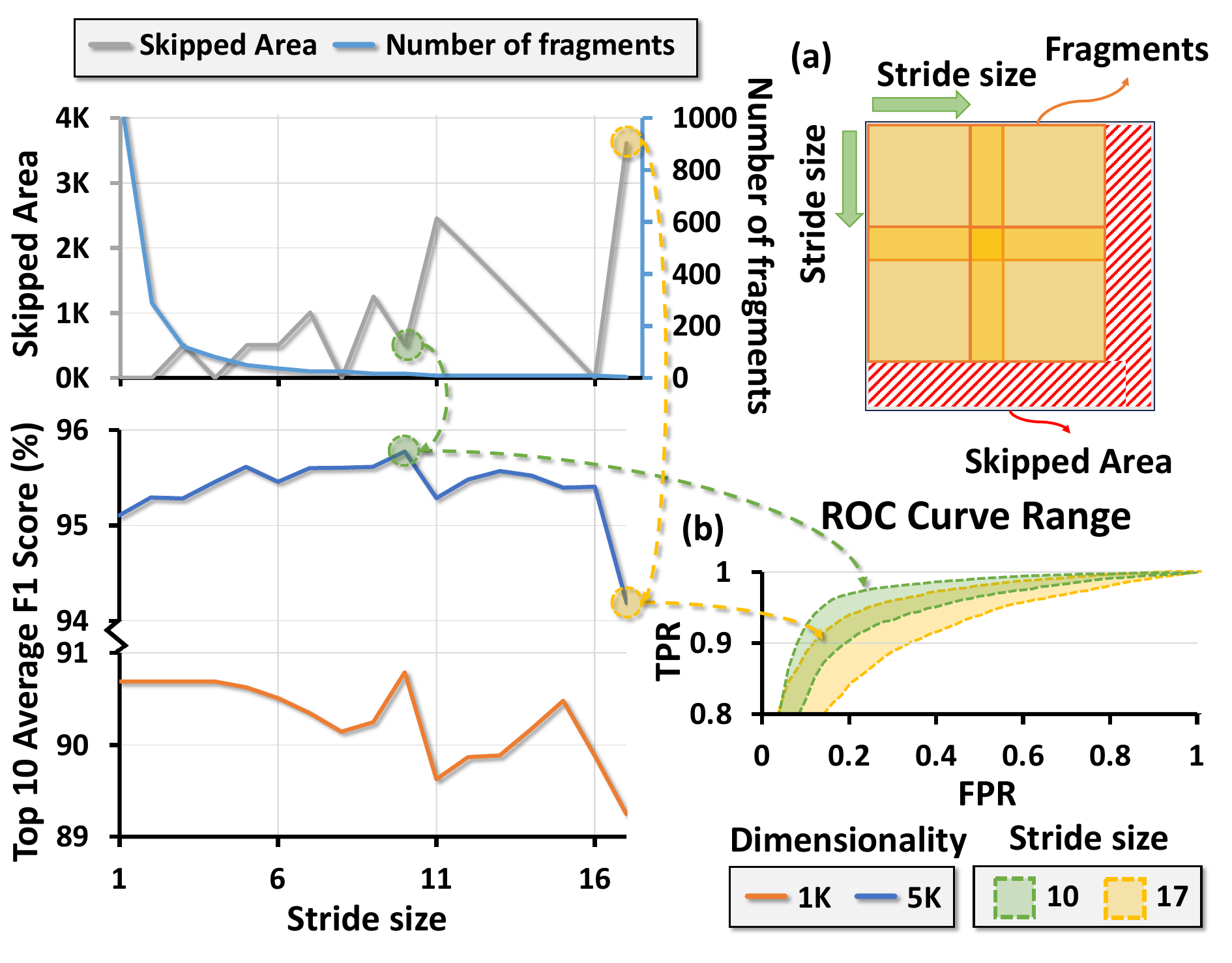}
    \vspace{-8mm}
    \caption{Effect of different stride sizes in terms of performance and computational amount when fragment size is 96}
    \label{fig:stridesize_test}
    \vspace{-3mm}
\end{figure}

\subsubsection{Fragment model performance}

In our initial set of experiments, we sought to compare the performance of an HDC model with a 10K dimensionality against other widely-used lightweight models as baselines, namely a small multi-layer perceptron (MLP) model and YOLOv4 tiny, in the context of object detection for the \emph{Fragment model}. The \emph{Fragment model} here refers to a specific configuration with a fragment size of 128.
\autoref{fig:fragment_model_compare} depicts ROC curves obtained from these different models when applied to the \emph{Fragment model}. Notably, on the left side of the ROC curves, even though the YOLOv4 tiny model has more than 5 million parameters which is much larger compared to other models making it hard to efficiently implement in near-sensor, it is evident that the YOLOv4 tiny model exhibits the lowest quality in terms of ROC curve performance. We assume it is due to the YOLOv4 tiny model's lack of performance on radar data as we can see in the previous work which reported the lowest performance of the YOLOv4 tiny model in terms of F1 score while having the minimum level of model size and latency~\cite{bochkovskiy2020yolov4}. Conversely, on the right side of the ROC curves, we observe the performance distinction between HDC and MLP in the TPR range of 0.8 to 1. Remarkably, HDC showcases the most discerning ROC curve compared to the other models. The quantified comparison of the Area Under the Curve (AUC) for the right ROC curves in \autoref{fig:fragment_model_compare} as presented in \autoref{tab:auc_compare} further confirms the highest performance of HDC compared to the baselines.

\begin{figure}
    \centering
    \includegraphics[width=1\linewidth]{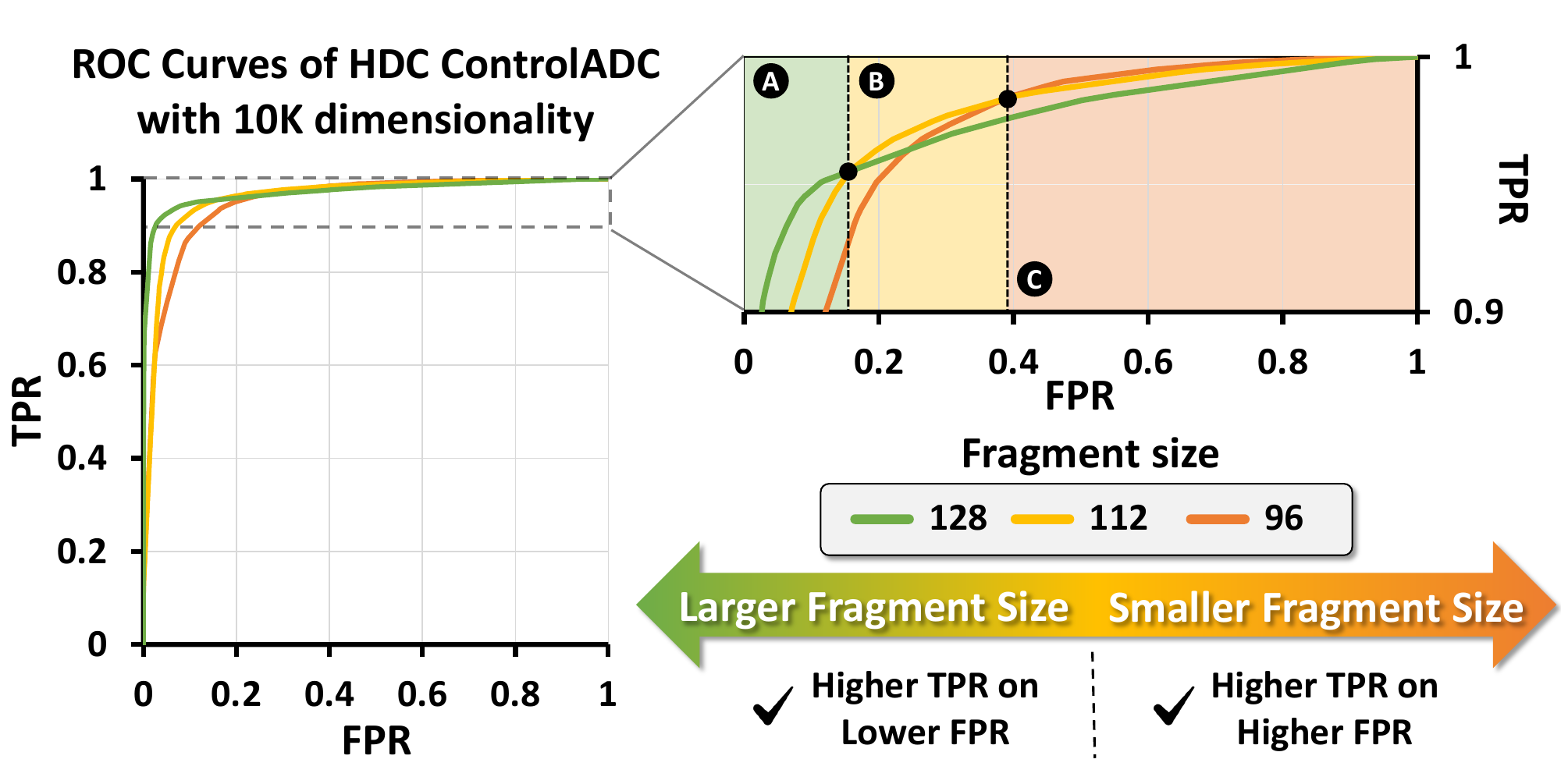}
    \vspace{-9mm}
    \caption{ROC Curve of HDC HyperSense with the dimensionality of 10K. As fragment size gets larger, we get a higher true positive rate on a higher false positive rate.}
    \label{fig:ROCCurves_by_winsize}
    \vspace{-5mm}
\end{figure}

\begin{figure}
    \centering
    \includegraphics[width=1\linewidth]{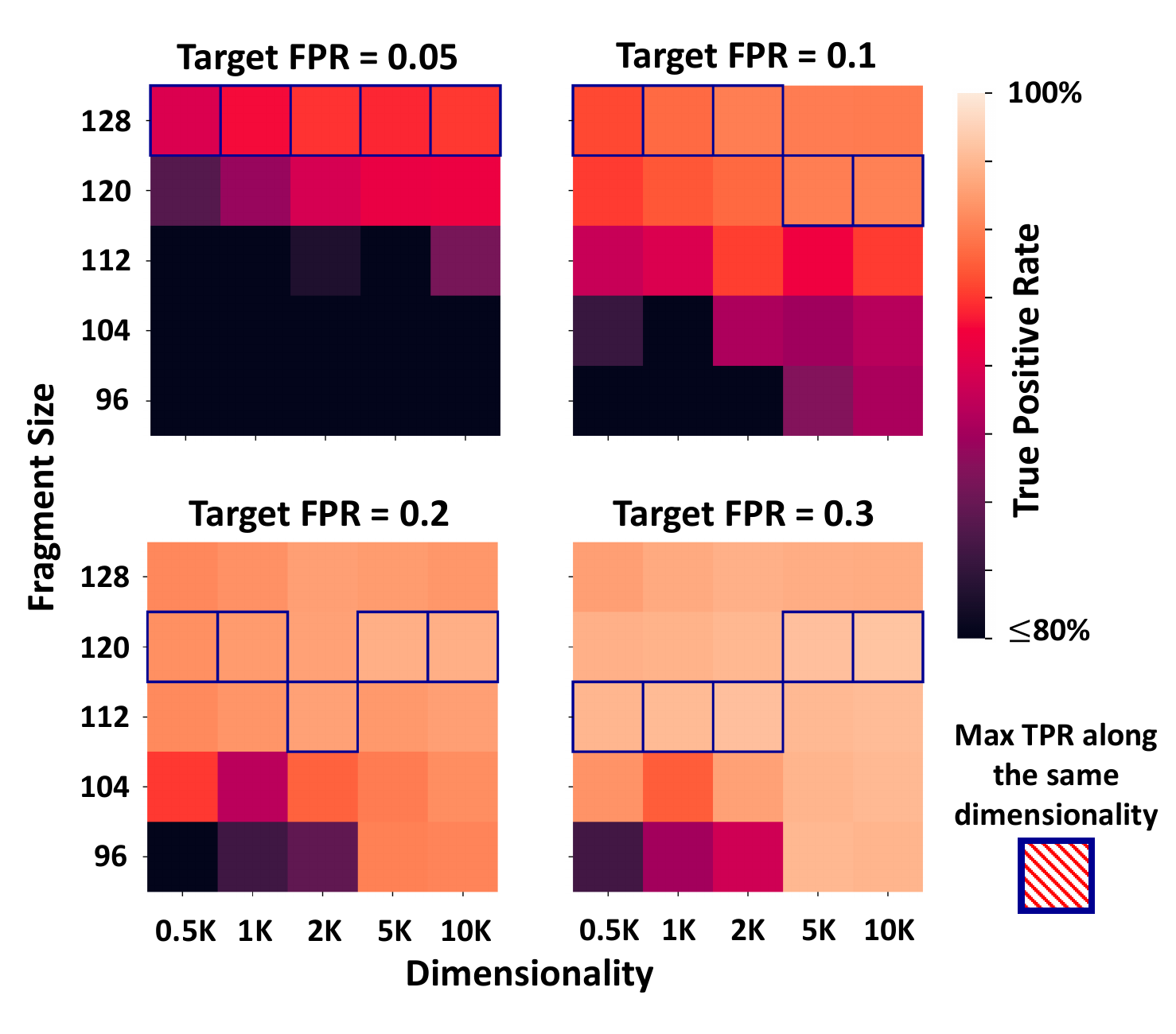}
    \vspace{-10mm}
    \caption{Exploration of maximum true positive rate (TPR) when targetting certain false positive rate (FPR) with different fragment sizes and dimensionalities.}
    \label{fig:winsize_dim_heatmaps}
    \vspace{-5mm}
\end{figure}

\subsubsection{Hyperparameters of HyperSense model exploration}

The \emph{HyperSense} model is composed of the \emph{Fragment model} and three essential hyperparameters: $T_{score}$, $T_{detection}$, and $stride$ as explained in \autoref{sec:HD_object}.
For the \emph{Fragment model} with a fixed fragment size of 128, which aligns with the sensing frames in the CRUW dataset, only one fragment is generated, resulting in a single ROC curve based on a single prediction score. This behavior is evident in \autoref{fig:fragment_model_compare}. However, when the \emph{Fragment model} employs a different fragment size, the \emph{HyperSense} model produces multiple prediction scores, necessitating the use of $T_{detection}$. Additionally, the \emph{HyperSense} model now processes a given frame of sensing data using a moving window approach, introducing another hyperparameter, $stride$, into the mix. Consequently, to comprehensively analyze the impact of each hyperparameter on the ROC curve, we undertook an exploration of different hyperparameter values.

In \autoref{fig:thresholds_test}, we present the results of exploring two thresholds, $T_{score}$ and $T_{detection}$, on two distinct \emph{Fragment model}s with varying sizes. As depicted in the left heatmaps, different values of $T_{detection}$ yield varying F1 scores across different $T_{score}$ values. This observation indicates that different selections of $T_{detection}$ give rise to distinct ROC curves. As a result, on the right side of the figure, we observe that the \emph{HyperSense} model now exhibits a range of ROC curves rather than a single ROC curve. 
This observation indicates the importance of selecting the appropriate ROC curve by identifying the highest TPR at a specific FPR, necessitating the right choice of $T_{detection}$ value.

The experimental results on different $stride$ sizes reveal that a larger $stride$ leads to a substantial skipping area (i.e., the parts of the fragment that do not fit into the frame), as illustrated in \autoref{fig:stridesize_test}.(a). 
Consequently, if an object is located within this skipped area, the model might miss detecting it, resulting in potential mispredictions. 
This effect is evident in the line graphs in \autoref{fig:stridesize_test}. The models with fewer skipped areas exhibit higher performance, while those with more skipped areas show lower performance as shown in the relationship between the skipped area line graph and the top 10 average F1 score graphs.
The ROC curves (shown in \autoref{fig:stridesize_test}.(b)) also support our observation: smaller $stride$ benefit the model.
Nonetheless, using a smaller $stride$ results in the generation of more fragments, which in turn increases the computational load. Hence, striking a balance between computation and performance becomes crucial, and the objective is to select the largest $stride$ that provides comparable performance to the model using a $stride$ of 1.

\begin{figure*}
    \centering
    \includegraphics[width=1\linewidth]{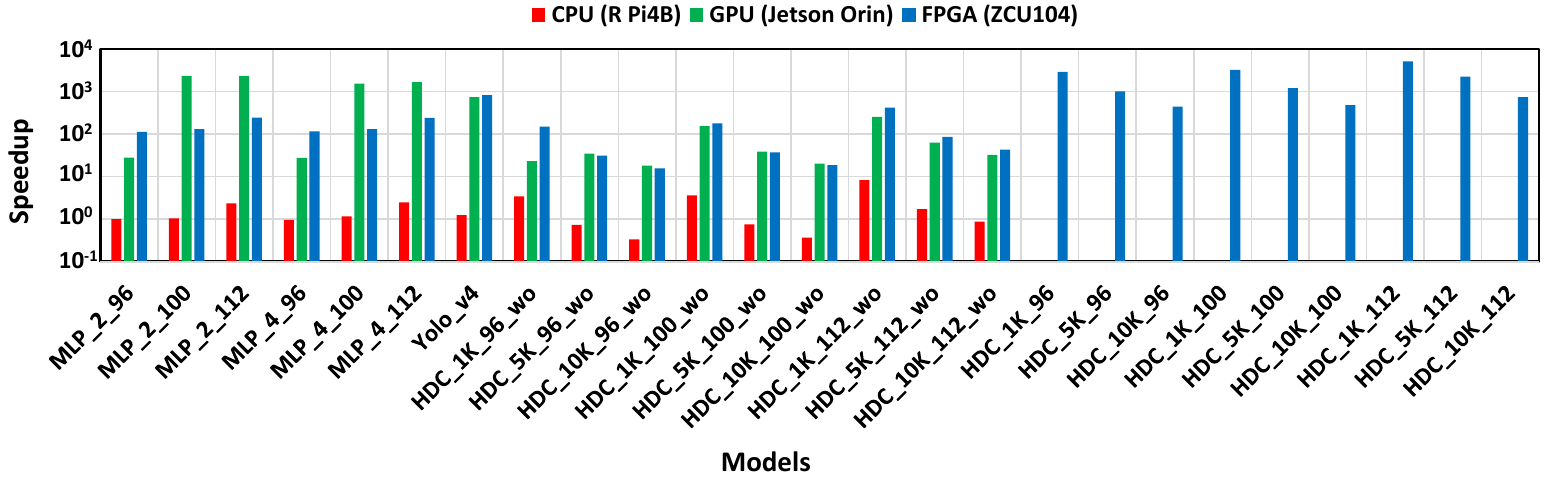}
    \vspace{-8mm}
    \caption{Cross models and cross platforms comparison. All three hardware platforms' operating power is between 5W to 15W. Here we suppose MLP with fragment size of 96 on R Pi 4B as the baseline.}
    \label{fig:res_speedup}
    \vspace{-5mm}
\end{figure*}

Interestingly, even though a stride size of 10 generates a larger skipped area compared to most stride sizes less than 10, its performance remains similar or even higher than the others. Considering the reduced number of fragments generated by larger stride sizes, leading to a possible decrease in computation, selecting a higher stride size that offers comparable performance to a stride size of 1 would be an efficient choice in a trade-off relationship.

\subsubsection{Fragment size effects on ROC curve}

From the preceding experiments, we observed that selecting appropriate hyperparameters can result in a \emph{HyperSense} model with a higher TPR. In this experiment, our focus shifted to the effect of fragment size on performance, considering a ROC curve composed of the highest TPR achievable from a given \emph{Fragment model} with a specific fragment size. 
The behavior of the ROC curve with different fragment sizes is illustrated in \autoref{fig:ROCCurves_by_winsize}.

At the lower FPR range \myCircled{A}, the \emph{HyperSense} model with the largest fragment size of 128 exhibits the highest TPR. However, as we move to the range of higher FPR, \emph{HyperSense} models with smaller fragment sizes achieve higher FPR. Specifically, in the middle range \myCircled{B}, a fragment size of 112 yields the highest TPR, while the smallest fragment size of 96 performs best in the highest FPR range \myCircled{C}. From this observation, we can capture a trend where in larger fragment sizes tend to perform better at lower FPRs, maximizing the TPR while minimizing quality loss due to reduced data granularity. Conversely, smaller fragment sizes are more effective at higher FPRs, where the risk of false positives is more manageable. This indicates that the choice of fragment size varies with a trade-off trend depending on the desired FPR.

Considering the results obtained with a dimensionality of 10K, we decided to conduct a comprehensive exploration of different fragment sizes and various dimensionalities to validate the observations from \autoref{fig:ROCCurves_by_winsize}. The heatmaps shown in \autoref{fig:winsize_dim_heatmaps} represent the true positive rates under the different target false positive rates. When targeting the lowest FPR of 0.05, the highest fragment size consistently exhibits the highest TPR for all dimensionalities. However, as we increase the target FPR, the fragment size that achieves the maximum TPR starts to decrease for all dimensionalities, aligning with the trend observed in \autoref{fig:ROCCurves_by_winsize}. These results further support the notion that the choice of fragment size plays a crucial role in achieving optimal performance in the \emph{HyperSense} model, depending on the desired trade-off between true positive and false positive rates.

\subsection{FPGA Resource Utilization}

\begin{table}[]
\caption{FPGA Resource Utilization on Xilinx ZCU104. The Hypervector Dimension is 5K and Fragment Size is 96. }
\label{tab:fpga_resource}
\vspace{-2mm}
\centering
\resizebox{0.4\textwidth}{!}{%
\begin{tabular}{c|cccc}
\toprule
              & LUT    & FF                        & BRAM   & DSP    \\ \midrule
Available     & 230K   & 460.8K                    & 312    & 1728   \\ \midrule
SA IP         & 137.5K & 40.8K                     & 216    & 256    \\
Classifier IP & 40.5K  & \multicolumn{1}{l}{50.1K} & 0      & 128    \\
Others        & 5.6K   & 41.1K                     & 16     & 0      \\ \midrule
Total         & 183.6K & 132K                      & 232    & 384    \\
Percentage    & 79.8\% & 28.6\%                    & 74.3\% & 22.2\% \\ \midrule
Frequency     & \multicolumn{4}{c}{100MHz}                           \\
Power         & \multicolumn{4}{c}{8.2W}                             \\ \bottomrule
\end{tabular}%
}
\vspace{-5mm}
\end{table}

In~\autoref{tab:fpga_resource}, we present the FPGA resource utilization of HDC accelerator on ZCU 104. We supposed the dimensionality of hypervector is 5K and the fragment size is 96. The data precision of each hypervector element is 8 bits. The encoding operation happens inside SA IP. Our classifier IP refers to classic HDC classifier FPGA accelerator design~\cite{imani2021revisiting}. We also include other peripheral IPs' resource utilization in~\autoref{tab:fpga_resource}, i.e., encoder IP~\cite{imani2021revisiting}, AXI Interconnect, and DRAM controller. The operating frequency of the accelerator is 100MHz with a power consumption of 8.2W. We got the power result based on Xilinx Power Estimator~\cite{becker2003power}.
For a single fragment, the HDC encoding and classification process takes 9397 clock cycles.
Although the operation on a single fragment does not show an obvious advantage in our system, the throughput of a whole radar image sensing shows a notable improvement using our tailored computation reuse scheme and pipeline data flow. The details about the computation reuse and pipeline execution flow are elaborated in~\ref{sec:FPGA acc}

\subsection{FPGA Accelerator Performance}\label{sec:FPGA acc}
\autoref{fig:res_speedup} presents the cross-platforms and cross-models comparison. For the hardware side, we picked three different platforms that operate at 10W levels, including Raspberry PI 4B CPU (R Pi 4B), NVIDIA Jetson Orin (Jetson Orin), and Xilinx ZCU 104. The operating power of R Pi 4B and Jetson Orin is 15W. We restricted all platforms' power consumption below 50\% of the sensor's power. For the model side, we compared MLP, YOLOv4, and HDC models. Each model is implemented on all three platforms. In~\autoref{fig:res_speedup}, we annotate MLP\_i\_j as MLP model with $i$ layers and fragment size $j$. We also annotate HDC\_d\_j as HDC models with hypervector dimension $d$ and fragment size $j$. For MLP and Yolo model acceleration on FPGA, we chose to use Xilinx DPU soft-core IP~\cite{kathail2020xilinx}. For HDC model acceleration on FPGA without computation reuse (HDC\_d\_\_wo), we implemented it based on the previous HDC FPGA framework~\cite{imani2021revisiting}. As is shown in~\autoref{fig:res_speedup}, after introducing HDC encoding computation reuse, HyperSense achieves on average 2.4$\times$ speedup when compared with MLP models running on Jetson Orin. When comparing with YOLOv4 running on Jetson Orin, HyperSense achieves on average 5.6$\times$ speedup. The average detection throughput of HyperSense on FPGA is around 303frames per second (FPS).

\subsection{End-to-end System Evaluation}

\begin{figure}
    \centering
    \includegraphics[width=1\linewidth]{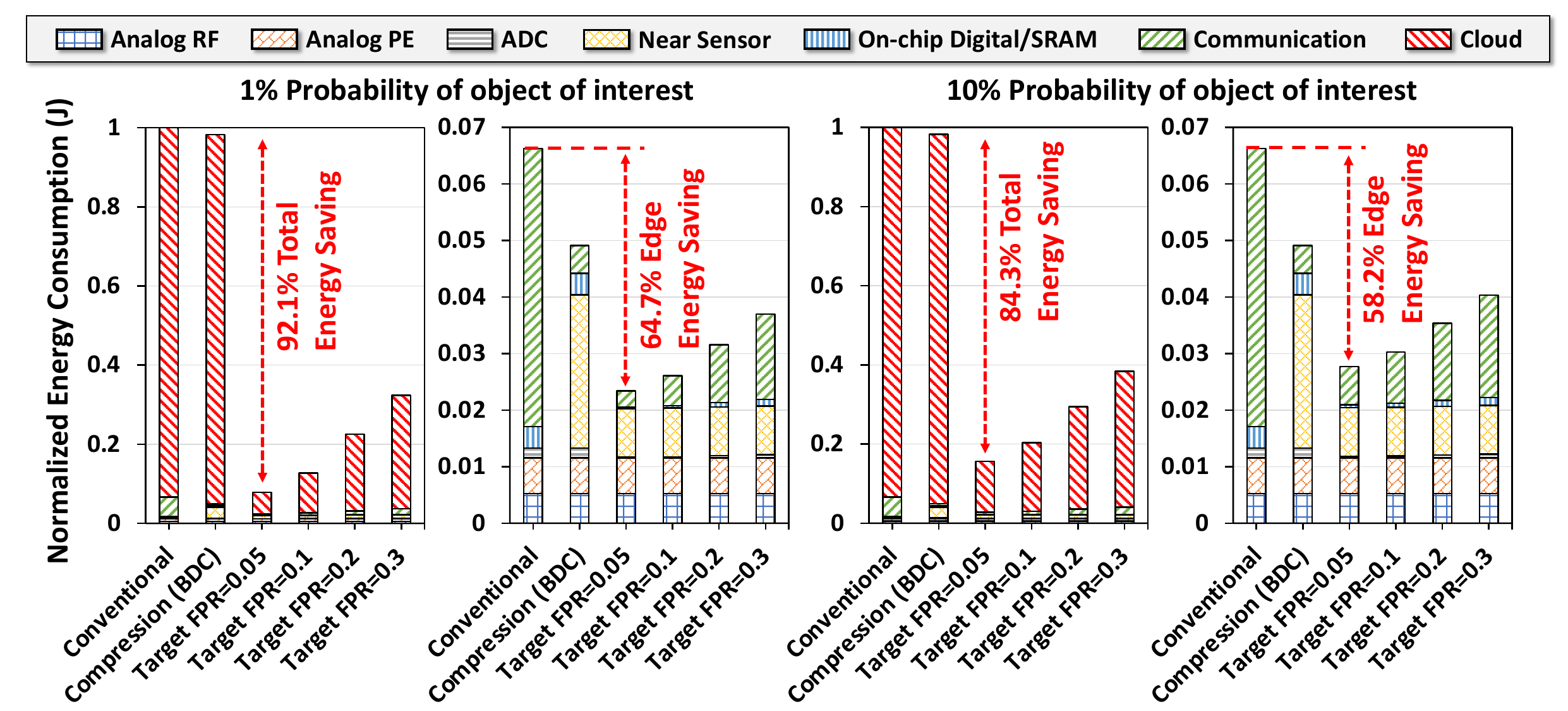}
    \vspace{-7mm}
    \caption{Energy consumption breakdown estimation in two different scenarios where the object of interest is infrequent (left) and 10 times frequent (right) for the conventional method, compressive sensing method, and ours with different target FPRs.}
    \label{fig:energybreakdown}
    \vspace{-6mm}
\end{figure}

\begin{table}[]
\caption{End-to-end evaluation of trade-off relationship between energy saving and quality loss of our framework when the probability of object of interest is 1\%.}
\label{tab:tradeoff}
\vspace{-2mm}
\centering
\resizebox{0.45\textwidth}{!}{%
\begin{tabular}{c|cccc}
\toprule
        Target FPR      & 0.05    & 0.1                        & 0.2  & 0.3    \\ \midrule
        Total Energy Saving      & \textbf{92.1\%}    & 89.8\%                        & 80.6\%   & 71.3\% \\
        Edge Energy Saving      & \textbf{64.7\%}    & 60.6\%                        & 52.4\%   & 44.2\% \\
        Quality Loss      & 7.44\%    & 4.93\%                        & 2.92\%   & \textbf{1.95\%} \\ \bottomrule
\end{tabular}%
}
\vspace{-5mm}
\end{table}

To assess the efficacy of our framework in conserving overall system energy, we conducted a comprehensive end-to-end system evaluation. Our analysis focused on estimating the average energy consumption for a single radar frame in a scenario where radar data is captured using a TI AWR1843 sensor and transmitted through a 3G network communication channel to a cloud server equipped with a resource-intensive machine-learning model for complex tasks. The energy cost estimation for the cloud server aligns with prior methodologies~\cite{selvan2023operating}.

In \autoref{fig:energybreakdown}, we present a breakdown of energy consumption in two scenarios: one with infrequent occurrences of the object of interest (1\% probability) and another with a tenfold increase in frequency. The comparison involves three methods: 1) the conventional method absensing an additional mechanism for energy conservation, 2) a widely recognized approach—compressive sensing, leveraging data compression, particularly Bit Depth Compression (BDC), a recent real-time application~\cite{hwang2023lossless}, and 3) our framework, which incorporates varying target FPRs. Notably, our approach demonstrates a noteworthy capability to conserve energy in both scenarios, affecting not only the total system but also the energy consumption at the edge by showing up to 92.1\% and 64.7\% energy saving respectively.

It is worth noting that the most energy-efficient case, targeting an FPR of 0.05, leads to the highest quality loss. This loss indicates that a portion of the data containing objects of interest, which was supposed to be transmitted to the cloud, was retained due to mispredictions by our near-sensor model, as elucidated in \autoref{tab:tradeoff}. Despite this trade-off relationship, our approach maintains a reasonable level of quality loss at approximately 1.95\%, while achieving more than a 71\% reduction in total system energy consumption.

\section{Discussion}
\subsection{Adapting \emph{HyperSense} to various sensors}

The adaptability of the \emph{HyperSense} system to various sensors, such as radar, cameras, and microphones, is fundamental to its broad utility across diverse applications. This system employs HDC to uniformly process segmented data units like frames, demonstrating that as long as the input can be transformed into a two-dimensional data unit, our framework can be applied, regardless of the sensor type. For example, in the case of microphones, a fast Fourier transform can convert audio into a two-dimensional spectrogram of fixed length. Our system's ability to interpret input data in standardized units (frames or fixed-length spectrograms) enables seamless integration with different sensor technologies. Additionally, the inherent robustness of the HDC architecture against noise and data quality variance, which are prevalent issues in sensors like microphones and LiDAR, further enhances the system's adaptability. This versatility not only expands the potential applications of our system but also bolsters its practicality for real-world deployments across various industries that require intelligent sensing solutions.

\subsection{Scalability and deployment challenges}
When integrating \emph{HyperSense} with other sensors, such as camera sensors, to perform object detection tasks on different types of data (such as images), two challenges may arise. The first challenge involves using the HDC model for complex vision tasks. Previous studies have indicated the necessity of integrating a Deep Neural Network (DNN) model with the HDC model~\mbox{\cite{hersche2022constrained,yun2024spatial}}. This integration introduces significant computational overheads, as DNN’s convolution operations are generally computation-intensive. For near-sensor processing tasks aiming for real-time image frame processing speeds, one potential solution is to add a domain-specific DNN accelerator. This accelerator would enhance the speed of DNN convolution operations~\mbox{\cite{lee2023comprehensive,dutta2022hdnn}}. It would be integrated with the HDC accelerator module via an on-chip connection protocol such as AMBA AXI~\mbox{\cite{sarekokku2012design,ni2023brain,chen2023hypergraf}}. 

The second potential challenge arises when using \emph{HyperSense} to process high-resolution data. To achieve real-time processing speeds with high-resolution input data, the most straightforward design choice is to increase the number of computing units in the accelerator to enhance computational parallelism. However, more computing units typically result in higher power consumption and a larger chip area. The near-sensor processing environment often faces strict power and space restrictions. Therefore, increasing computational parallelism to handle high-resolution input data may cause the whole system to fail to fit into the near-sensor environment. One potential solution is to integrate a variational autoencoder (VAE) to compress the high-resolution input data~\mbox{\cite{vahdat2020nvae}}.

\section{Conclusions}
We introduce \emph{HyperSense}, an innovative co-designed hardware and software that tries to solve the gap between intelligent sensing and machine learning. Our cutting-edge system effectively manages data generation from the Analog-to-Digital Converter (ADC) modules by predicting object presence in sensor data. Our framework outperforms other models with the highest Area Under the Curve (AUC). At the same time, the FPGA-based hardware exhibits remarkable speedups, rendering \emph{HyperSense} an ideal solution for real-time intelligent sensing and data processing by showing up to 92.1\% energy saving on end-to-end system evaluation.

\section*{Acknowledgements}
This work was supported in part by the DARPA Young Faculty Award, National Science Foundation \#2127780, \#2319198, \#2321840, \#2312517, Semiconductor Research Corporation (SRC), Office of Naval Research Young Investigator Program Award, grants \#N00014-21-1-2225 and \#N00014-22-1-2067, the Air Force Office of Scientific Research under award \#FA9550-22-1-0253, and generous gifts from Xilinx and Cisco.

\bibliographystyle{IEEEtranS}
\bibliography{refs}

\end{document}